\documentclass[sigconf,nonacm]{acmart}
\usepackage{pifont}
\usepackage{subfig}
\usepackage{multirow}
\usepackage{pythonhighlight}
\usepackage{threeparttable}
\usepackage{bm}
\usepackage{colortbl}
\usepackage{xcolor}
\usepackage{hyperref}
\definecolor{lightgray}{rgb}{0.94,0.94,0.94}
\definecolor{colorl}{rgb}{1,0.88,0.70}
\AtBeginDocument{%
  }

\setcopyright{acmlicensed}
\copyrightyear{2018}
\acmYear{2018}
\acmDOI{XXXXXXX.XXXXXXX}
\acmConference[Conference acronym 'XX]{Make sure to enter the correct
  conference title from your rights confirmation email}{June 03--05,
  2018}{Woodstock, NY}
\acmISBN{978-1-4503-XXXX-X/2018/06}

\acmSubmissionID{312}

\begin{document}

\title{BiECVC: Gated Diversification of Bidirectional Contexts for Learned Video Compression}


\author{Wei Jiang}
\email{jiangwei.lvc@bytedance.com}
\orcid{0000-0001-9169-1924}
\affiliation{%
  \institution{Bytedance}
  \city{}
  \country{}
}
\author{Junru Li}
\orcid{0000-0001-7603-8599}
\email{lijunru@bytedance.com}
\affiliation{%
  \institution{Bytedance}
  \city{}
  \country{}
}
\author{Kai Zhang}
\orcid{0000-0002-6627-0009}
\email{zhangkai.video@bytedance.com}
\affiliation{%
  \institution{Bytedance}
  \city{}
  \country{}
}
\author{Li Zhang}
\orcid{0000-0003-2118-4876}
\email{lizhang.idm@bytedance.com}
\affiliation{%
  \institution{Bytedance}
  \city{}
  \country{}
}

\renewcommand{\shortauthors}{Jiang, et al.}

\begin{abstract}
Recent forward prediction-based learned video compression (LVC) methods have 
achieved impressive results, even surpassing VVC reference software VTM under the Low Delay B (LDB) configuration.
In contrast, learned bidirectional video compression (BVC) remains 
underexplored and still lags behind its forward-only counterparts. 
This performance gap is mainly due to the limited ability to extract diverse and accurate contexts: most existing BVCs primarily exploit temporal motion 
while neglecting non-local correlations across frames. 
Moreover, they lack the adaptability to dynamically suppress harmful contexts arising from 
fast motion or occlusion.
To tackle these challenges, we propose BiECVC, a BVC framework that 
incorporates diversified local and non-local context modeling along with adaptive context gating.
For local context enhancement, BiECVC reuses high-quality features from lower layers and aligns them using decoded motion vectors without introducing extra motion overhead.
To model non-local dependencies efficiently, we adopt a linear attention mechanism that balances performance and complexity.
To further mitigate the impact of inaccurate context prediction, we introduce Bidirectional Context Gating, inspired by data-dependent decay in recent autoregressive language models, 
to dynamically filter contextual information based on conditional coding results.
Extensive experiments demonstrate that BiECVC achieves state-of-the-art performance, reducing the bit-rate by 13.4\% and 15.7\% 
compared to VTM 13.2 under the Random Access (RA) configuration with intra periods of 32 and 64, respectively.
To our knowledge, BiECVC is the first learned video codec to surpass VTM 13.2 RA across all standard test datasets.
\end{abstract}




\maketitle

\section{Introduction}
Video compression aims to represent visual signals using fewer bits while maintaining 
high reconstruction quality. Traditional video coding methods, such as 
VVC~\cite{bross2021overview}, have been developed over decades and 
have achieved impressive performance. However, their progress is fundamentally 
constrained by the hand-crafted design of coding tools, leading to a performance plateau.
In contrast, learned video compression (LVC)~\cite{lu2019dvc,li2021deep,jiang2025ecvc,jiang2024lvc,vctmentzer2022,hu2021fvc,guo2023learning,yang2023insights,rippel2021elf,liu2020learned,lin2022dmvc} 
is optimized in a fully end-to-end manner, offering greater flexibility and the potential for superior performance. 
As a result, it has garnered increasing attention in recent years.\par
Current state-of-the-art (SOTA) LVCs~\cite{li2022hybrid,li2023neural,li2024neural,qilong2024,Qi_2023_CVPR,jia2025practical,jiang2025ecvc} follow the conditional coding paradigm~\cite{li2021deep}, where
temporal conetxts\footnote[1]{Contexts refer to predictions used for computing residuals or conducting conditional coding, which differ from the contexts used in entropy models.} are used as priors for a reduction in conditional entropy. 
Existing approaches~\cite{lu2019dvc,li2021deep,li2022hybrid,li2023neural,li2024neural,jia2025practical,jiang2025ecvc,jiang2024lvc} 
primarily focus on forward prediction, in which each frame (P-frame) can 
only reference frames from earlier time steps. 
Although this strategy has led to impressive progress, with recent methods
~\cite{li2023neural, li2024neural, jiang2025ecvc} even surpassing VVC reference software VTM under the Low 
Delay B (LDB) configuration, bidirectional prediction, 
where a frame (B-frame) can reference both past and future frames, 
remains relatively underexplored.
In theory, bidirectional prediction offers richer contextual information and holds greater potential for compression efficiency.
However, current bidirectional video compression (BVC) methods~\cite{zhai2025llbvc,sheng2025bi,chen2024bcanf,alexandre2023hierarchical} 
still underperform compared to their forward-only counterparts~\cite{li2024neural,li2023neural,qilong2024,jiang2025ecvc}.
Such performance gap is largely due to the limited ability of existing methods to fully exploit bidirectional temporal priors,
as well as their insufficient adaptability in handling complex motion and occlusions.\par
To better understand the limitations of existing BVCs~\cite{sheng2025bi,chen2024bcanf,zhai2025llbvc,alexandre2023hierarchical,yang2021learning,yang2020learning},
we analyze the factors contributing to the superior performance of traditional codecs such 
as VVC under bidirectional prediction.
First, VVC supports the selection of up to $4\sim 6$ reference frames~\cite{girod2000efficiency,flierl2001multihypothesis,flierl2002rate}, 
offering significantly more diverse temporal information compared to current 
BVCs. Most existing BVCs utilize only a single forward and 
a single backward frame as references, which may fall short in modeling complex 
motion dynamics or handling occluded regions. Second, in VVC,
motion estimation is based on minimizing Sum of Absolute Differences (SAD)~\cite{vanne2006high}, which captures block-level similarity.
However, small SAD values do not necessarily imply real motion; 
they may simply indicate pixel fidelity. Thus, 
VVC can benefit from flexible content matching, even across 
frames with weak temporal relevance, a capability largely absent in current optical flow-based BVCs.
Finally, VVC assigns adaptive weights~\cite{chen2016generalized,winken2019weighted} to multiple reference frames through rate-distortion (RD) 
optimization, enabling flexible utilization of contextual information. In contrast, 
existing learned BVC approaches typically lack mechanisms for context-wise weighting, 
making it difficult to selectively enhance informative cues while suppressing noisy 
or misleading signals.\par
To address above issues, we propose bidirectional context diversification for 
enhanced exploitation of temporal priors, and bidirectional context gating to adaptively weight contexts based on conditional coding results.
We categorize bidirectional contexts into local and non-local types~\cite{jiang2025ecvc}.
Local contexts capture correlations with explicit motion, 
while non-local contexts correspond to similar regions without clear motion trajectories as illustrated in Figure~\ref{fig:nlc}.\par
To diversify local contexts, we incorporate additional reference frames from lower hierarchical layers.
In BVCs, such lower-layer frames are generally assigned higher bit-rates and therefore offer better reconstruction quality, 
making them valuable sources for reliable motion-aligned context. 
To align these frames with the current target frame without incurring additional motion coding overhead, we reuse the decoded motion vectors already estimated for primary reference frames.
\par
To diversify non-local contexts, we adopt a linear attention mechanism~\cite{jiang2025ecvc,shen2021efficient} 
to capture non-local correlations across frames. Linear attention compares each element 
in the current frame with all elements in reference frames, based on learned 
attention-driven similarity.
We note that the process of aggregating contexts using attention scores 
fundamentally aligns with the principle of block matching in VVC, which selects 
reference candidates by minimizing the SAD. However, unlike hand-crafted SAD, our approach enables flexible
similarity optimization in an end-to-end manner.\par
In bidirectional context gating, we draw inspiration from data-dependent 
decay in linear-time autoregressive language modeling~\cite{yang2024gated,qin2023hierarchically}.
To compute the gating matrix, we first concatenate the 
latent features with the corresponding contexts, and apply a nonlinear transform. 
The resulting gating matrix is used to modulate 
the magnitude of each context feature.
Interestingly, this nonlinear transform of the latent feature and contexts 
can be interpreted as an attempt of conditional coding—performed 
without weighting—to assess their relevance. The output of this initial interaction 
serves as guidance for the actual weighted conditional coding.\par
Based on proposed techniques, we propose a bidirectional video compression model with Gated Diversification of Bidirectional Contexts, termed BiECVC.
To accelerate BiECVC, to our knowledge, we introduce the \textit{first} feature cache mechanism for BVCs to efficiently reuse pre-computed features.
Experiments demonstrate that BiECVC achieves SOTA performance, reducing bit-rate by 13.4\% and 15.7\%
 compared to the VTM 13.2 RA with intra periods of 32 and 64.
To the best of our knowledge, BiECVC is the first learned video codec to outperform VTM 13.2 RA across all standard test datasets.
Our contributions are summarized as follows:
\begin{itemize}
  \item To diversify local contexts, BiECVC exploits high-quality features from lower layers and aligns them using decoded motion vectors, without additional motion overhead.
  \item To diversify non-local contexts, linear attention mechanism is employed to 
  capture correlations across frames with unclear motion trajectories based on learned attention-driven
  similarity. 
  \item To enhance context weighting, we introduce a bidirectional context
  gating mechanism that dynamically emphasizes beneficial information while suppressing harmful noise, drawing inspiration from
  recent advancements in linear-time autoregressive language modeling.
  \item Experiments demonstrate that our BiECVC achieves SOTA performance,
  reducing bit-rate by 13.4\% and 15.7\% over VTM 13.2 RA with intra periods of 32 and 64.
  To our knowledge, BiECVC is the first learned video codec to surpass VTM 13.2 RA across all standard test datasets.
\end{itemize}
\section{Related Works}
\subsection{Learned Video Compression with Forward Prediction}
Most existing learned video compression models focus on forward prediction.
DVC~\cite{lu2019dvc} is one of the pioneering frameworks in this domain,
following the residual coding paradigm. It employs an optical flow network~\cite{ranjan2017optical} for motion estimation,
compressing both motion and residuals between the original and predicted frames using neural networks.
To enhance performance, the DVC framework has been further refined with various techniques,
including resolution-adaptive coding~\cite{hu2020improving}, scale-space flow warping~\cite{agustsson2020scale},
selective compression~\cite{shi2022alphavc}, and coarse-to-fine prediction~\cite{hu2022coarse}.\par
Despite advancements in residual coding-based methods, they fail to fully exploit temporal priors, 
as their performance is fundamentally constrained by the entropy of the residuals~\cite{li2021deep}.
To address this issue, Li~\textit{et al.}~\cite{li2021deep} propose the conditional coding paradigm,
where the neural network learns the correlations between the current frame and temporal priors automatically,
rather than relying on predefined residuals, leading to improved compression efficiency.
This paradigm is further refined with temporal context mining~\cite{sheng2022temporal} for better temporal dependency exploitation.
Regarding entropy modeling, Li~\textit{et al.} propose hybrid spatial-temporal coding~\cite{li2022hybrid} and 
quadtree-based entropy models~\cite{li2023neural} to minimize the bit-rate of latent representations. 
Additionally, hierarchical quality and offset diversity~\cite{chan2022basicvsr++} are leveraged to enhance temporal adaptability.
To mitigate accumulated errors in long coding chains, Li~\textit{et al.}~\cite{li2024neural} 
introduce temporal-propagated context refreshment, significantly improving long-term performance.
Recently, Jiang~\textit{et al.}~\cite{jiang2025ecvc} propose ECVC, which captures long-range dependencies across multiple frames. 
With an improved training strategy, ECVC achieves state-of-the-art performance in low-delay configurations.
\begin{figure*}
  \includegraphics[width=\textwidth]{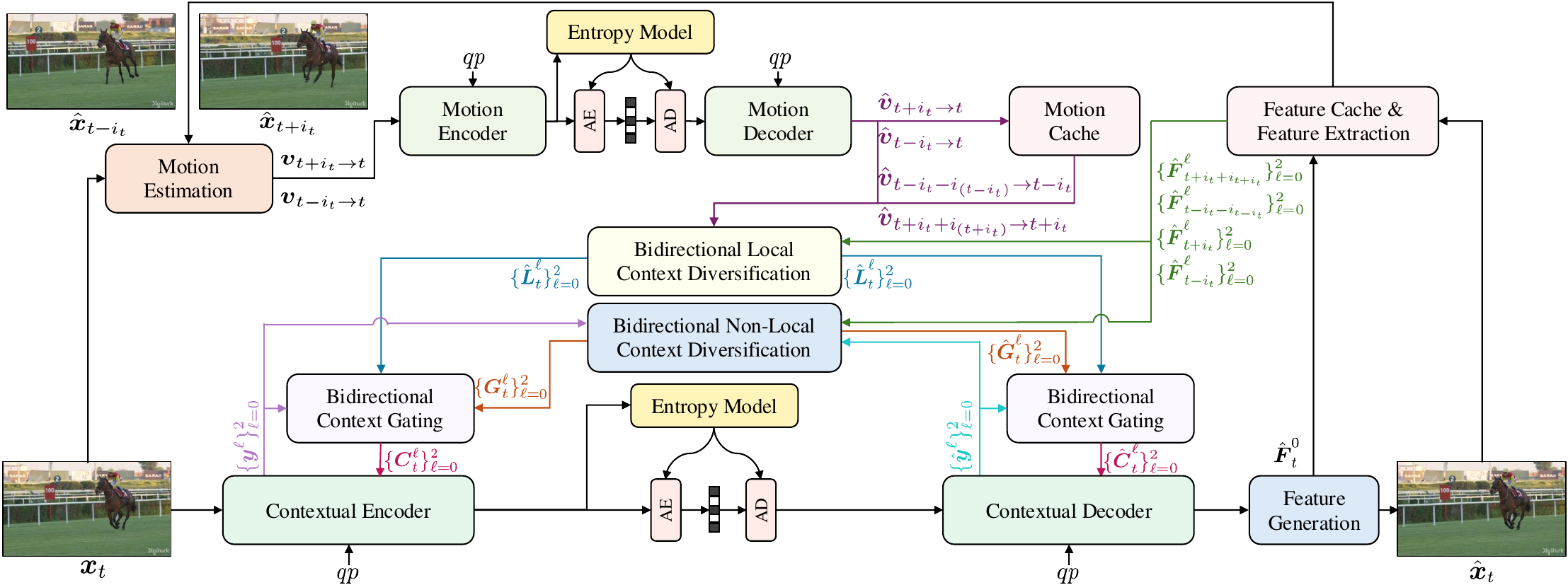}
  \caption{Overall framework of BiECVC. The quantization parameter (\textit{qp}) controls variable-rate compression. 
  Local contexts $\{\hat{\boldsymbol{L}}^{\ell}_t\}_{\ell=0}^2$ are shared between the encoder and decoder. 
  The non-local contexts, $\{{\boldsymbol{G}}^{\ell}_t\}_{\ell=0}^2$ and $\{\hat{\boldsymbol{G}}^{\ell}_t\}_{\ell=0}^2$, 
  are extracted separately at the encoder and decoder. 
  The final contexts, ${\boldsymbol{C}_t^{\ell}}$ and ${\hat{\boldsymbol{C}}_t^{\ell}}$, 
  are used at the contextual encoder and decoder, respectively.
  AE and AD are arithmetic encoding and decoding, respectively.}
  \label{fig:overall}
\end{figure*}
\subsection{Learned Video Compression with Bidirectional Prediction}
Compared to forward prediction, bidirectional prediction is more challenging and has been relatively unexplored in learned video compression.
Wu~\textit{et al.}~\cite{wu2018video} propose the first motion-free video compression framework, which relies on frame interpolation. 
The residual between the interpolated and original frames is compressed using a neural network.
This approach is further refined by Xu~\textit{et al.}~\cite{xu2024ibvc} by advanced interpolation techniques.
Djelouah~\textit{et al.}~\cite{djelouah2019neural} introduce residual coding in the latent space while
Alexandre~\textit{et al.}~\cite{alexandre2023hierarchical} propose a two-layer conditional augmented normalizing flow to improve expressiveness.\par
In contrast, most BVCs retain motion estimation and motion compensation modules.
Yang~\textit{et al.}~\cite{yang2020learning,yang2021learning} employ an optical flow network for bidirectional motion estimation, 
where both the predicted frame (obtained via flow warping) and the motion itself are compressed.
Pourreza~\textit{et al.}~\cite{pourreza2021extending} introduce an interpolation-based approach, 
generating a predicted frame by interpolating two warped reference frames before performing residual coding.
Yang~\textit{et al.}~\cite{yang2024ucvc} develop UCVC, a conditional coding-based method capable of compressing both P-frames 
and B-frames within a single model.
Sheng~\textit{et al.} propose DCVC-B~\cite{sheng2025bi}, a bidirectional video compression model that extends temporal propagation 
and multi-scale contexts from forward prediction to bidirectional prediction.\par
Despite advancements in bidirectional video compression, existing models still 
lag behind their forward-only counterparts. 
This performance gap stems from challenges such as limited contextual information
and errors of predicted contexts. 
Current methods typically use flow maps for motion compensation and conditional coding, 
which overlooks non-local correlations without explicit movements, leading to limited context diversity.
Moreover, predicted errors in contexts caused by large motions and non-overlapped regions introduce extra noise,
reducing coding efficiency.
\begin{figure}
  \includegraphics[width=0.485\textwidth]{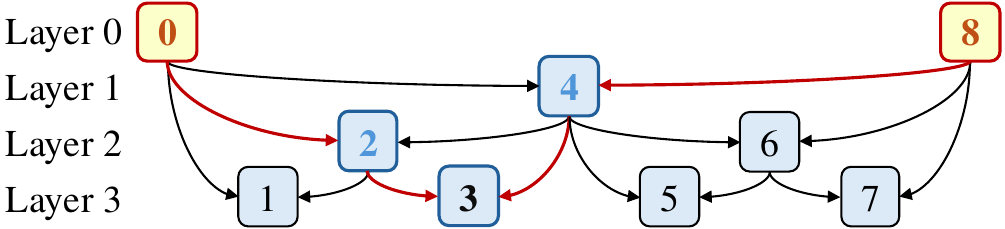}
  \caption{Coding Structure when Intra Period is 8. "Yellow" denotes I-frames, "Blue" denotes B-frames.
  When coding 3-th frame, \{0,2,4,8\}-th frames serve as references.}
  \label{fig:gop}
\end{figure}
\begin{figure*}
  \includegraphics[width=\textwidth]{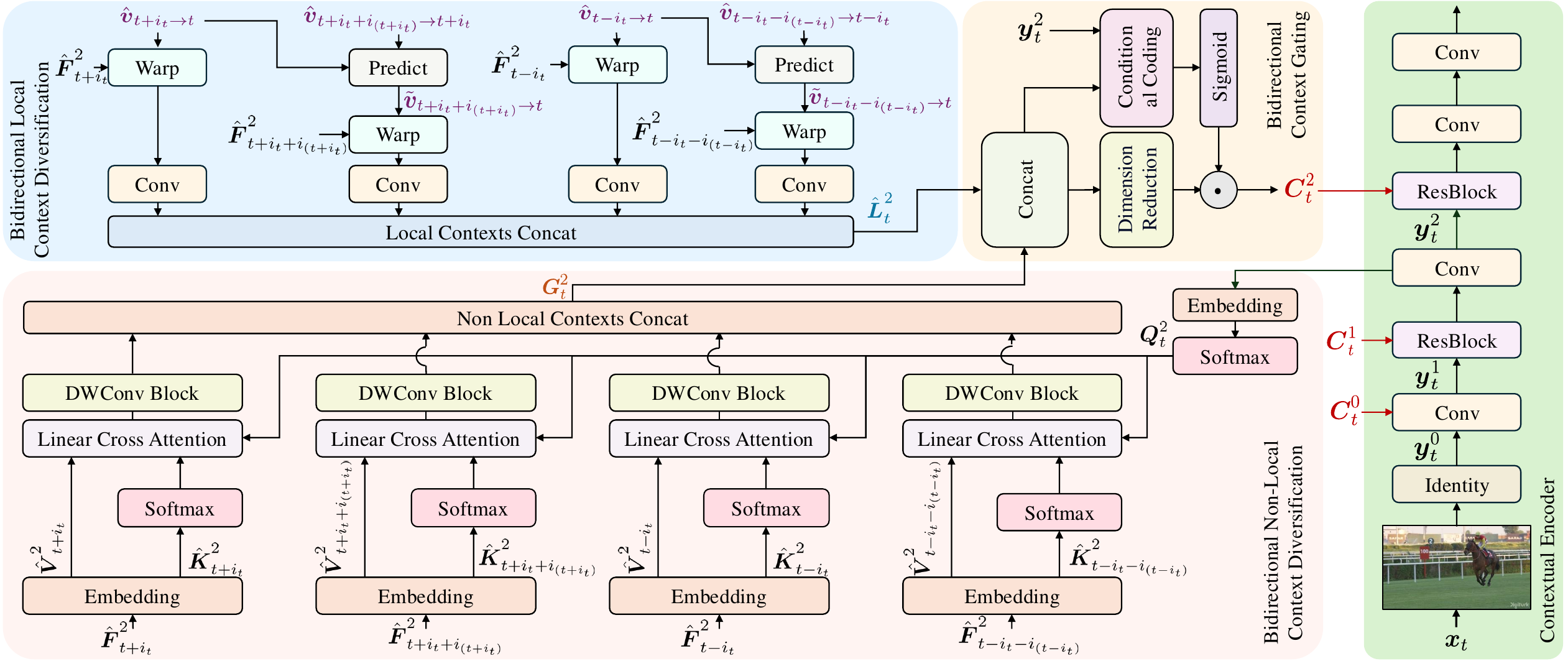}
  \caption{Illustrations of the proposed Bidirectional Local Context Diversification (BLCD) and Bidirectional Non-Local Context Diversification (BNLCD)
  and Bidirectional Context Gating (BCG). The process is conditional coding $\boldsymbol{y}_t^2$ at the encoder side.
  The decoding process mirrors the encoding process.}
  \label{fig:bcd}
\end{figure*}
\section{Method}
\subsection{Overview}
The overall architecture of our proposed BiECVC is presented in Figure~\ref{fig:overall} and Figure~\ref{fig:bcd}.
Our BiECVC is based on ECVC~\cite{jiang2025ecvc} and the offset diversity~\cite{chan2022basicvsr++} module is removed for lower complexity.
\textit{In Section~\ref{sec:blcd}, \ref{sec:bnlcd} and \ref{sec:BCG}, we mainly take the encoder side as example.
The process of the decoder side mirrors that of the encoder side}.
\subsubsection{Background}
Given sequence $\{\boldsymbol{x}_t\}_{t=0}^{N-1}\in \mathbb{R}^{3\times H\times W}$, where $t$ is the time step,
$N$ is the number of frames, $H$ and $W$ are the height and width of a frame respectively,
bidirectional video compression aims to efficiently encode the sequence
by dividing it into multiple group of pictures (GOP) and make the reconstructions $\{\hat{\boldsymbol{x}}_t\}_{t=0}^{N-1}$
with high quality. Within each GOP, Intra (I) frames are encoded independently,
while Bidirectional (B) frames are compressed using contexts from both backward and forward references.\par
In the bidirectional coding structure, frames within an GOP are organized into multiple hierarchical layers.
For example, when the IP is $8$, the frames are divided into 4 layers as illustrated in Figure~\ref{fig:gop}.
Lower layers are assigned more bits and higher 
quality because they are referenced by multiple frames at higher layers.\par
BiECVC employs the feature propagation mechanism in DCVC series~\cite{sheng2022temporal} and ECVC~\cite{jiang2025ecvc}.
Specifically, let $i_t$ represent the temporal interval between the current frame $\boldsymbol{x}_t$ and 
its two reference frames $\hat{\boldsymbol{x}}_{t-i_t}$ and $\hat{\boldsymbol{x}}_{t+i_t}$.
During encoding,
$\boldsymbol{x}_t$ is first transformed to multi-scale $\{\boldsymbol{y}_t^\ell\}_{\ell=0}^2 \in \mathbb{R}^{C^{\ell} \times \frac{H}{2^\ell} \times \frac{W}{2^\ell}}$.
Propagated features 
$\{\hat{\boldsymbol{F}}_{t-i_t}^0,\hat{\boldsymbol{F}}_{t+i_t}^0\}\in \mathbb{R}^{D^0\times H\times W}$ from lower layers
and their downsampled
$\{\hat{\boldsymbol{F}}_{t-i_t}^{\ell}\}_{\ell=1}^2 \in \mathbb{R}^{D^{\ell}\times \frac{H}{2^{\ell}}\times \frac{W}{2^{\ell}}}$ 
and $\{\hat{\boldsymbol{F}}_{t+i_t}^{\ell}\}_{\ell=1}^2 \in \mathbb{R}^{D^{\ell}\times \frac{H}{2^{\ell}}\times \frac{W}{2^{\ell}}}$ 
are warped by motion vectors $\hat{\boldsymbol{v}}_{t-i_t\rightarrow t}$ and $\hat{\boldsymbol{v}}_{t+i_t\rightarrow t}$ 
to serve as the multi-scale temporal priors for conditional coding corresponding $\{\boldsymbol{y}_t^\ell\}_{\ell=0}^2$.
\begin{figure}
  \includegraphics[width=0.485 \textwidth]{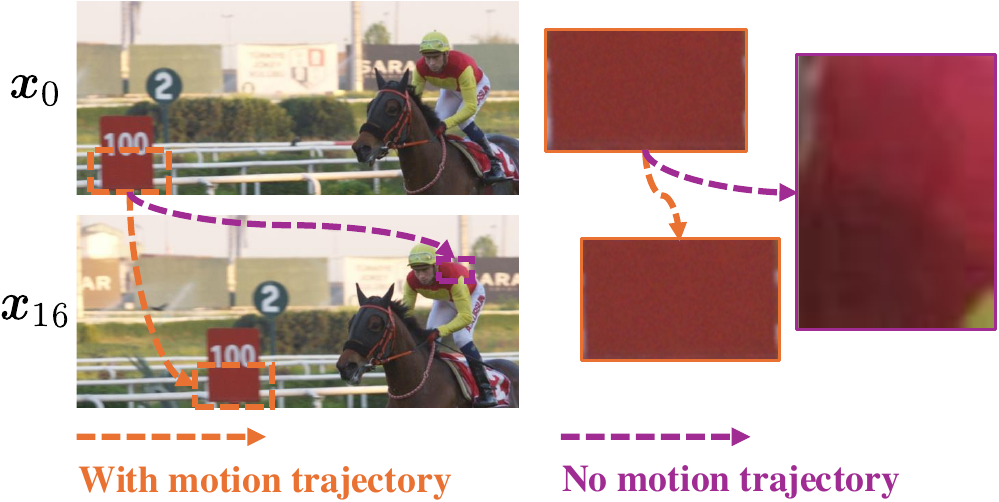}
  \caption{Visualization of local context (with motion trajectory) and non-local context (no motion trajectory).}
  \label{fig:nlc}
\end{figure}
\subsubsection{Motion Estimation and Compression}
To leverage temporal priors,
we employ SPyNet~\cite{ranjan2017optical} to estimate the motion vectors
$\{\boldsymbol{v}_{t-i_t\rightarrow t},\boldsymbol{v}_{t+i_t\rightarrow t}\}\in \mathbb{R}^{2\times H\times W}$,
which represent the motion from time step $t-i_t$ to $t$ and from $t+i_t$ to $t$, respectively.
To efficiently compress the motion vectors, $\boldsymbol{v}_{t-i_t\rightarrow t}$ and $\boldsymbol{v}_{t+i_t\rightarrow t}$
are concatenated along the channel dimension and fed into a motion encoder.
On the decoder side, the reconstructed tensor with four channels is split along the channel dimension
to obtain the reconstructed motion vectors
$\{\hat{\boldsymbol{v}}_{t-i_t\rightarrow t},\hat{\boldsymbol{v}}_{t+i_t\rightarrow t}\}\in \mathbb{R}^{2\times H\times W}$.
\subsubsection{Bidirectional Contextual Compression}
\par
In BiECVC, we propose the Bidirectional Local Context Diversification (BLCD) and Bidirectional Non-Local Context Diversification (BNLCD)
to diversify the temporal contexts.
\textit{Local contexts are defined as regions exhibiting explicit motion, 
whereas non-local contexts refer to similar regions that do not display explicit movement as illustrated in Figure~\ref{fig:nlc}.}
For diverse local contexts,
BLCD incorporates additional local contexts from time steps $t-i_t-i_{(t-i_t)}$ and $t+i_t+i_{(t+i_t)}$ 
in addition to local contexts at $t-i_t$ and $t+i_t$
for better quality of lower layers.
For example, in Figure~\ref{fig:gop}, when $t=3$, contexts from $t=\{0,8\}$ are employed in addition to contexts from $t=\{2,4\}$.
For effective non-local context modeling,
BNLCD employs linear attention mechanisms~\cite{jiang2025ecvc,jiang2023mlic,jiang2023mlicpp,shen2021efficient,jiang2025mlicv2} to capture
non-local contexts at time steps $t-i_t$, $t+i_t$, $t-i_t-i_{(t-i_t)}$, and $t+i_t+i_{(t+i_t)}$ with linear complexity.
Considering the \textit{non-overlapped regions} between temporal 
contexts and the current feature due to fast motions, and accumulated errors during feature propagation, 
inspired by the data-dependent decay mechanism in recent linear attentions for 
autoregressive language modeling~\cite{yang2024gated,qin2023hierarchically,qinhgrn2,zhang2024gated,sun2023retentive,yang2024gated2}, 
we propose a Conditional Coding-Aware Bidirectional Context Gating (BCG) mechanism to adaptively adjust context weights via gating.
In BCG, we first attempt conditional coding $\{\boldsymbol{y}_t^\ell\}_{\ell=0}^2$ 
using the available contexts.
The results guide the generation of a weighting mask $\boldsymbol{M}$, which selectively adjusts the influence of different contexts
to suppress harmful errors and non-overlapped contexts.
In implementation, we introduce feature cache mechanism to reuse precomputed features
for acceleration.
\subsection{Bidirectional Local Context Diversification}
\label{sec:blcd}
In existing bidirectional video compression models~\cite{sheng2025bi,zhai2025llbvc,yang2024ucvc}, the conditional coding of $\boldsymbol{y}_t^{\ell}$ 
relies on motion compensation from two reference frames at time steps $t-i_t$ and $t+i_t$:
\begin{equation} 
  \hat{\boldsymbol{L}}_{t-i_t\rightarrow t}^{\ell} = \mathcal{W}(\hat{\boldsymbol{F}}_{t-i_t}^{\ell}, 
  \hat{\boldsymbol{v}}_{t-i_t\rightarrow t}), \hat{\boldsymbol{L}}_{t+i_t\rightarrow t}^{\ell} = 
  \mathcal{W}(\hat{\boldsymbol{F}}_{t+i_t}^{\ell}, \hat{\boldsymbol{v}}_{t+i_t\rightarrow t}), 
\end{equation}
where $\mathcal{W}$ represents warping with optical flow.
\par
However, in a hierarchical bidirectional coding structure, lower-layer reference frames/features generally 
have higher quality and lower reconstruction errors, which can improve compression efficiency. 
Additionally, videos often contain complex motion patterns, such as fast movements~\cite{fan2016motion}, 
affine transformations~\cite{zhang2018improved}, and occlusions~\cite{huang2006analysis},
using only two reference frames may not be sufficient to handle these challenging scenarios.
To address these issues, we propose the Bidirectional Local Context Diversification (BLCD), 
as illustrated in Figure~\ref{fig:bcd}. In BLCD, we leverage priors from time steps $t-i_t-i_{(t-i_t)}$ and $t+i_t+i_{(t+i_t)}$, 
as the corresponding features $\hat{\boldsymbol{F}}_{t-i_t-i_{(t-i_t)}}^{\ell}$ and $\hat{\boldsymbol{F}}_{t+i_t+i_{(t+i_t)}}^{\ell}$ 
are available in most cases during the coding of $\boldsymbol{y}_t^{\ell}$.
For example, in Figure~\ref{fig:gop}, when $t=3$, features from $t=\{0,8\}$ are employed in addition to features from $t=\{2,4\}$.
A key challenge in utilizing these additional contexts is how to warp the features at $t-i_t-i_{(t-i_t)}$ and $t+i_t+i_{(t+i_t)}$ 
to align with the current time step $t$. Direct motion estimation and compression would not only increase computational 
complexity but also require extra bits to encode motions.
Fortunately, if both $\hat{\boldsymbol{F}}_{t-i_t-i_{(t-i_t)}}^\ell$ and $\hat{\boldsymbol{F}}_{t-i_t}^\ell$ are available, 
the motion $\hat{\boldsymbol{v}}_{t-i_t-i_{(t-i_t)} \rightarrow t-i_t}$ has already been decoded. 
This allows us to \textit{predict} the motion $\tilde{\boldsymbol{v}}_{t-i_t-i_{(t-i_t)} \rightarrow t}$ by accumulating flows~\cite{wu2023accflow}.
Similarly, we could \textit{predict} the motion $\tilde{\boldsymbol{v}}_{t+i_t+i_{(t+i_t)} \rightarrow t}$ 
based on $\hat{\boldsymbol{v}}_{t+i_t+i_{(t+i_t)} \rightarrow t+i_t}$
and $\hat{\boldsymbol{v}}_{t+i_t\rightarrow t}$. For simplicity, we omit the convolutional layer in Figure~\ref{fig:bcd}.
Thus, the additional local contexts can be formulated as:
\begin{equation}
  \begin{aligned}
  \hat{\boldsymbol{L}}_{t-i_t-i_{(t-i_t)}\rightarrow t}^{\ell} &= \mathcal{W}(\hat{\boldsymbol{F}}_{t-i_t-i_{(t-i_t)}}^{\ell}, \tilde{\boldsymbol{v}}_{t-i_t-i_{(t-i_t)}\rightarrow t}),\\
  \hat{\boldsymbol{L}}_{t+i_t + i_{(t+i_t)}\rightarrow t}^{\ell} &= \mathcal{W}(\hat{\boldsymbol{F}}_{t+i_t+i_{(t+i_t)}}^{\ell}, \tilde{\boldsymbol{v}}_{t+i_t+i_{(t+i_t)}\rightarrow t}).
  \end{aligned}
\end{equation}
To balance reference diversity and computational cost, we set $\ell \in \{1,2\}$, avoiding the inclusion of $\ell=0$ 
for the local contexts from $t-i_t-i_{(t-i_t)}$ and $t+i_t+i_{(t+i_t)}$.
\begin{figure}
  \includegraphics[width=0.485\textwidth]{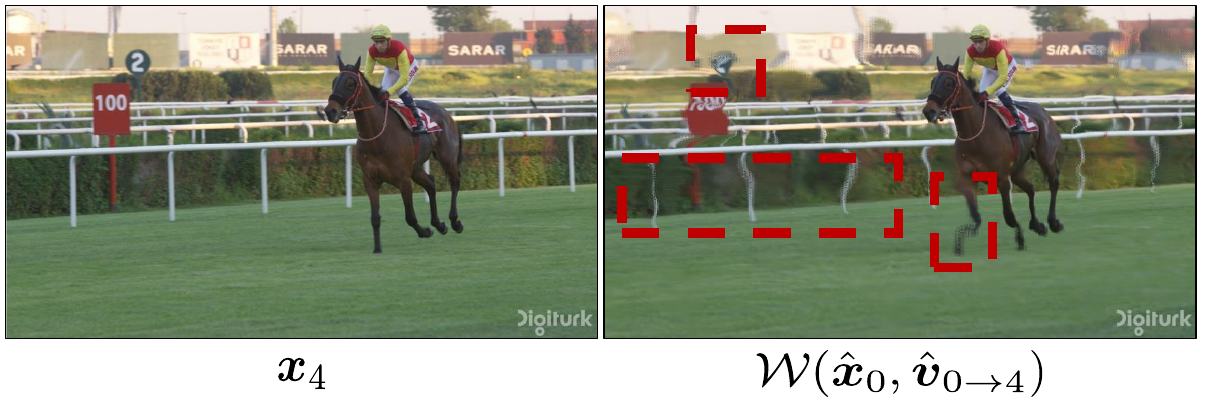}
  \caption{Visualization of predicted errors in optical flow-based warping.
  The frames are from "Jockey" sequence in UVG~\cite{mercat2020uvg} dataset.}
  \label{fig:warped}
\end{figure}
\subsection{Bidirectional Non-Local Context Diversification}
\label{sec:bnlcd}
To enhance contextual diversity, BiECVC introduces Bidirectional Non-Local Context 
Diversification (BNLCD) to capture non-local correlations as illustrated in Figure~\ref{fig:nlc}. 
This requires comparing all elements of the current feature $\boldsymbol{y}_t^\ell$ with those of the reference features, necessitating a non-local receptive field.
Given the flexibility of conditional coding, we follow ECVC~\cite{jiang2025ecvc} 
and adopt linear attention~\cite{shen2021efficient}, allowing the network to learn 
correlations automatically. For high-resolution features, linear attention is 
more efficient and memory-friendly than vanilla attention~\cite{vaswani2017attention}, as it avoids the 
explicit computation of large attention maps with quadratic complexity. Additionally, 
since non-local attention directly models feature similarities, explicit warping of features is unnecessary.
To maintain a balanced parameter count, we utilize a shared linear attention layer across all available features.
First, current $\boldsymbol{y}_t^\ell$ is embedded to a query $\boldsymbol{Q}_t^{\ell} \in \mathbb{R}^{D^{\ell}\times \frac{H}{2^\ell}\times \frac{W}{2^\ell}}$. 
The available features $\hat{\boldsymbol{F}}_{t-i_t}^{\ell}$ and $\hat{\boldsymbol{F}}_{t+i_t}^{\ell}$ are embedded 
into keys $\{\hat{\boldsymbol{K}}_{t-i_t}^\ell, \hat{\boldsymbol{K}}_{t+i_t}^\ell\} \in \mathbb{R}^{D^{\ell}\times \frac{H}{2^\ell}\times \frac{W}{2^\ell}}$ 
and values $\{\hat{\boldsymbol{V}}_{t-i_t}^\ell, \hat{\boldsymbol{V}}_{t+i_t}^\ell\}\in \mathbb{R}^{D^{\ell}\times \frac{H}{2^\ell}\times \frac{W}{2^\ell}}$.
For simplicity, we omit the DWConv Block~\cite{jiang2024llic} in Figure~\ref{fig:bcd}.
The process is formulated as:
\begin{equation}
  \begin{bmatrix}\boldsymbol{G}_{t- i_t\rightarrow t}^\ell \\ \boldsymbol{G}_{t+ i_t\rightarrow t}^\ell\end{bmatrix} = 
  \underbrace{\textrm{SM}_2(\boldsymbol{Q}_t^{\ell})
    \overbrace{\begin{bmatrix}
      \textrm{SM}_1(\hat{\boldsymbol{K}}_{t-i_t}^\ell)^{\top}\hat{\boldsymbol{V}}_{t-i_t}^\ell\\
      \textrm{SM}_1(\hat{\boldsymbol{K}}_{t+i_t}^\ell)^{\top}\hat{\boldsymbol{V}}_{t+i_t}^\ell
    \end{bmatrix}}^{\mathcal{O}\left((D^{\ell})^2HW\right)}}_{\mathcal{O}\bigl((D^{\ell})^2HW+(D^{\ell})^2HW\bigr)}.
\end{equation}
This formulation applies two independent softmax (SM) operations on row and column—allowing 
the key-value product to be computed first, reducing complexity to linear order. 
Additionally, each row of the implicit similarity matrix 
(\textit{e.g.}, $\textrm{softmax}_2(\boldsymbol{Q}_t^{\ell})\textrm{softmax}1(\hat{\boldsymbol{K}}_{t-i_t}^\ell)^{\top}$) 
sums to 1, ensuring a normalized attention distribution across all positions~\cite{jiang2025ecvc,shen2021efficient}.\par
For more diverse non-local contexts, we explore the non-local contexts from time step $t-i_t-i_{(t-i_t)}$ and $t+i_t+i_{(t+i_t)}$.
To balance reference diversity and complexity, we set $\ell \in \{1,2\}$. The process is formulated as:
\begin{equation}
  \begin{aligned}
    \begin{bmatrix}
  \boldsymbol{G}_{t-i_t-i_{(t-i_t)}\rightarrow t} ^\ell\\
   \boldsymbol{G}_{t+i_t+i_{(t+i_t)}\rightarrow t}^\ell\end{bmatrix} = 
  \textrm{SM}_2(\boldsymbol{Q}_t^{\ell})
   {\begin{bmatrix}
      \textrm{SM}_1(\hat{\boldsymbol{K}}_{t-i_t-i_{(t-i_t)}}^\ell)^{\top}\hat{\boldsymbol{V}}_{t-i_t-i_{(t-i_t)}}^\ell\\
      \textrm{SM}_1(\hat{\boldsymbol{K}}_{t+i_t+i_{(t+i_t)}}^\ell)^{\top}\hat{\boldsymbol{V}}_{t+i_t+i_{(t+i_t)}}^\ell
    \end{bmatrix}}.
  \end{aligned}
\end{equation}
\subsection{Conditional Coding Aware Bidirectional Context Gating}
\label{sec:BCG}
In bidirectional video compression, effectively utilizing the captured local contexts
$\hat{\boldsymbol{L}}_t^{\ell}=\{\hat{\boldsymbol{L}}_{t-i_t-i_{(t-i_t)}\rightarrow t}^\ell$, 
$\hat{\boldsymbol{L}}_{t-i_t\rightarrow t}^{\ell}$, $\hat{\boldsymbol{L}}_{t+i_t\rightarrow t}^{\ell}$,
$\hat{\boldsymbol{L}}_{t+i_t+i_{(t+i_t)}\rightarrow t}^{\ell}\}$,
and non-local contexts $\boldsymbol{G}_t^{\ell}=\{{\boldsymbol{G}}_{t-i_t-i_{(t-i_t)}\rightarrow t}^{\ell}$, 
${\boldsymbol{G}}_{t-i_t\rightarrow t}^{\ell}$, 
$\hat{\boldsymbol{G}}_{t+i_t\rightarrow t}^{\ell}$,
${\boldsymbol{G}}_{t+i_t-i_{(t+i_t)}\rightarrow t}^{\ell}\}$,
are crucial for achieving high compression efficiency.
However, as illustrated in Figure~\ref{fig:warped}, the predicted local contexts may contain errors due to non-overlapping regions or 
large motion. While conditional coding is highly flexible, fixed weights may fail to mitigate these errors. 
Therefore, for regions with unreliable contexts, 
it is essential to suppress their influence and favor intra coding (\textit{i.e.}, coding without external contexts).
Moreover, due to feature propagation, prediction errors may accumulate across layers, leading to performance degradation.\par
Recent advancements in linear-time autoregressive language modeling~\cite{yang2024gated,yang2024gated2} have demonstrated that 
\textit{data-dependent decay} plays a crucial role in preserving important information while suppressing irrelevant details. 
Formally, this process is defined as:
\begin{figure*}
  \includegraphics[width=\textwidth]{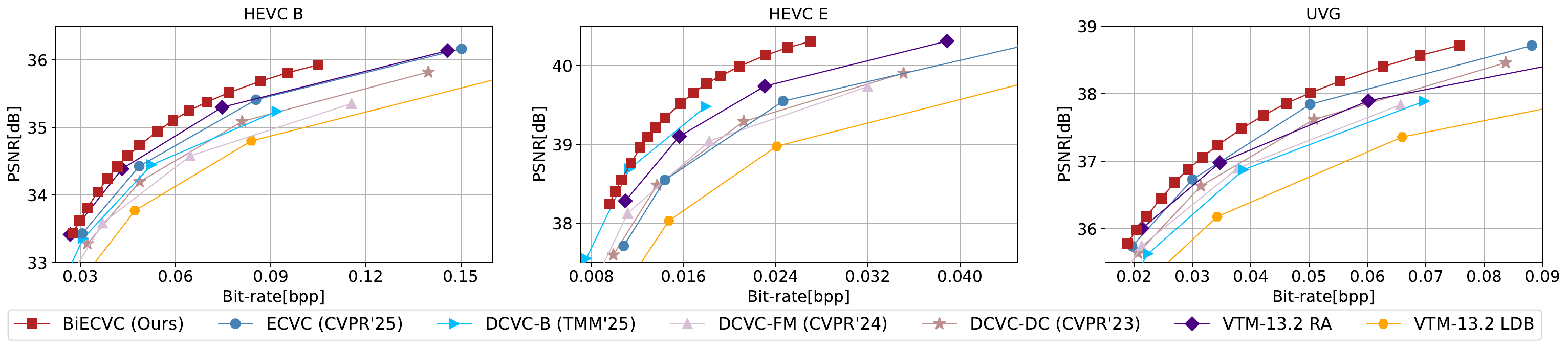}
  \caption{Rate-Distortion curves on HEVC B, HEVC E, and UVG dataset. The intra period is 32 with 96 frames.}
  \label{fig:rd}
\end{figure*}
\begin{equation} 
  \boldsymbol{S}_t = \boldsymbol{S}_{t-1} \odot \boldsymbol{M}_{t} + \boldsymbol{k}_t^{\top}\boldsymbol{v}_t, \quad \boldsymbol{o}_t = \boldsymbol{q}_t \boldsymbol{S}_t, 
\end{equation}
where $\boldsymbol{S}_t$ represents the state at time $t$, $\boldsymbol{M}_{t}$ is the data-dependent gating matrix, and $\boldsymbol{q}_t, \boldsymbol{k}_t, \boldsymbol{v}_t$ correspond to queries, keys, and values, respectively. The output is denoted as $\boldsymbol{o}_t$.
Drawing an analogy to video compression, we interpret $\hat{\boldsymbol{x}}_t$ as the output and $\hat{\boldsymbol{F}}_t$ as the current state. The process can be abstracted as:
\begin{equation} 
  \hat{\boldsymbol{F}}_t^0 = \hat{\boldsymbol{y}}^0_t + {\mathcal{F}}(\underbrace{\hat{\boldsymbol{F}}^0_{t-i_t}, \hat{\boldsymbol{F}}^0_{t+i_t}, \dots}_{\textrm{Previous States}}), 
  \hat{\boldsymbol{x}_t} ={\boldsymbol{W}}\hat{\boldsymbol{F}}^0_t,
\end{equation}
where $\mathcal{F}$ represents operations such as warping and attention, $\boldsymbol{W}$ represents convolutional weights.\par
Inspired by this connection, we propose the Bidirectional Context Gating (BCG).
At the encoder side, we generate the data-dependent decay matrix $\boldsymbol{M}_t^\ell$ based 
on $\boldsymbol{y}_t^{\ell}$, $\hat{\boldsymbol{L}}_t^{\ell}$, 
and ${\boldsymbol{G}}_t^{\ell}$. To reduce computational complexity, a dimension reduction (DR) branch is introduced before gating.
\textit{The matrix generation follows the principle of conditional coding (CC), meaning the network effectively performs conditional coding once, 
and the results guide the actual conditional coding.}
At the encoder side, the process is:
\begin{equation}
  {\boldsymbol{M}}_t^\ell = \omega(\textrm{CC}(\boldsymbol{y}_t^{\ell}, \hat{\boldsymbol{L}}_t^{\ell}, {\boldsymbol{G}}_t^{\ell})), \boldsymbol{C}_t^{\ell} = \boldsymbol{M}_t^\ell \odot \textrm{DR}(\hat{\boldsymbol{L}}_t^{\ell}, {\boldsymbol{G}}_t^{\ell}),
\end{equation}
Similarly, at the decoder side, the process is:
\begin{equation}
  {\hat{\boldsymbol{M}}}_t^\ell = \omega(\textrm{CC}(\hat{\boldsymbol{y}}_t^{\ell}, \hat{\boldsymbol{L}}_t^{\ell}, {\hat{\boldsymbol{G}}}_t^{\ell})), \hat{\boldsymbol{C}}_t^{\ell} = \hat{\boldsymbol{M}}_t^\ell \odot \textrm{DR}(\hat{\boldsymbol{L}}_t^{\ell}, {\hat{\boldsymbol{G}}}_t^{\ell}),
\end{equation}
where $\omega$ denotes the Sigmoid function, following HGRN~\cite{qin2023hierarchically}.
At the $0$-th scale, the features propagated to the next layers and reconstruction are computed by Feature Generation (FG)~\cite{li2023neural} and convolutions:
\begin{equation}
 \hat{\boldsymbol{F}}_t^0 = \textrm{FG}(\hat{\boldsymbol{y}}_t^0, \hat{\boldsymbol{C}}_t^{0}), \hat{\boldsymbol{x}}_t = \boldsymbol{W}\hat{\boldsymbol{F}}_t^0.
\end{equation}
With the BCG module, the network can effectively suppress harmful errors while emphasizing beneficial contexts.
\subsection{Feature Cache for Acceleration}
In bidirectional video compression, a single frame may serve as a reference for multiple frames. For instance, in Figure~\ref{fig:gop}, 
$\hat{\boldsymbol{x}}_0$ is referenced by frames $\{\boldsymbol{x}_t\}_{t=1}^4$. This implies that the multi-scale features, keys, and 
values computed for $\hat{\boldsymbol{x}}_0$ 
in BLCD and BNLCD during the encoding of $\boldsymbol{x}_4$ can also be reused for encoding $\{\boldsymbol{x}_t\}_{t=1}^3$.
To efficiently manage these reusable features, to our knowledge, we introduce the \textit{ first feature cache} mechanism for BVCs. Features that will be referenced 
later are stored in the cache, while those no longer needed are removed to control memory usage. 
This caching strategy improves computational efficiency, leading to an inference speedup of up to $18\%$.
\begin{table*}  
  \centering
  \setlength{\tabcolsep}{2.2mm}{
  \small
  \renewcommand\arraystretch{1.1}
    \begin{threeparttable}{
  \begin{tabular}{@{}cccccccccccccccc@{}}
  \toprule
  \multicolumn{1}{c}{\multirow{2}{*}{Method}} & \multicolumn{1}{c}{\multirow{2}{*}{{Venue}}}              & \multicolumn{1}{c}{\multirow{2}{*}{{Type}}}    & \multicolumn{1}{c}{\multirow{1}{*}{{Intra}}} & \multicolumn{1}{c|}{\multirow{1}{*}{{Frame}}}           & \multicolumn{7}{c}{{BD-Rate (\%) w.r.t. VTM-13.2 RA~\cite{bross2021overview}}}  \\
  \multicolumn{1}{c}{}      & \multicolumn{1}{c}{}                           & \multicolumn{1}{c}{}        & \multicolumn{1}{c}{Period}        & \multicolumn{1}{c|}{Number}                   & \multicolumn{1}{c}{HEVC B}    & \multicolumn{1}{c}{HEVC C}  & \multicolumn{1}{c}{HEVC D} & \multicolumn{1}{c}{HEVC E}  & \multicolumn{1}{c}{UVG} & \multicolumn{1}{c}{MCL-JCV} & \multicolumn{1}{c}{Average} \\       \midrule                                      
  \multicolumn{1}{c}{VTM-13.2 LDB~\cite{bross2021overview}}     & \multicolumn{1}{c}{---}                  & \multicolumn{1}{c}{LD}                & 32 &  \multicolumn{1}{c|}{96}   & $51.6$   & ${37.3}$ & $38.6$ & $67.1$ & $50.6$ &{${40.6}$} & $47.6$\\  
  \rowcolor{lightgray}\multicolumn{1}{c}{DCVC-TCM~\cite{sheng2022temporal}}     & \multicolumn{1}{c}{TMM'22}              & \multicolumn{1}{c}{LD}              & 32 &  \multicolumn{1}{c|}{96}    & 90.0 & 125.8 & 78.1 & 159.0& 64.9&77.7&99.3 \\
  \multicolumn{1}{c}{DCVC-HEM~\cite{li2022hybrid}}    & \multicolumn{1}{c}{ACMMM'22}                   & \multicolumn{1}{c}{LD}            & 32 &  \multicolumn{1}{c|}{96}      & 43.4& 61.3& 26.3 &74.8 &20.1& 30.9& 42.8 \\
  \rowcolor{lightgray}\multicolumn{1}{c}{DCVC-DC~\cite{li2023neural}}      & \multicolumn{1}{c}{CVPR'23}                   & \multicolumn{1}{c}{LD}            & 32 &  \multicolumn{1}{c|}{96}      & 25.2 &24.0 &-1.1 &21.5 &3.8 &12.2 &14.3 \\
  \multicolumn{1}{c}{DCVC-FM~\cite{li2024neural}}        & \multicolumn{1}{c}{CVPR'24}                 & \multicolumn{1}{c}{LD}           & 32 &  \multicolumn{1}{c|}{96}      &34.4& 29.9 &4.1& 20.7 &13.6 &23.4& 21.0\\
  \rowcolor{lightgray}\multicolumn{1}{c}{ECVC~\cite{jiang2025ecvc}}      & \multicolumn{1}{c}{CVPR'25} & \multicolumn{1}{c}{LD}  & 32 &  \multicolumn{1}{c|}{96}& 8.2 &10.7 &-12.5& 20.8& -8.2& 2.3& 3.6   \\
  \multicolumn{1}{c}{B-CANF~\cite{chen2024bcanf}}      & \multicolumn{1}{c}{TCSVT'23} & \multicolumn{1}{c}{RA} & 32 &  \multicolumn{1}{c|}{96} & 86.7 &102.7 &54.6 &89.3 &64.8 &79.7 &79.6\\
  \rowcolor{lightgray}\multicolumn{1}{c}{UCVC~\cite{yang2024ucvc}}      & \multicolumn{1}{c}{DCC'24} & \multicolumn{1}{c}{RA} & 32  & \multicolumn{1}{c|}{96}& 47.7 &46.5 &16.7 &29.8 &35.0 &64.1 &40.0\\
  \multicolumn{1}{c}{APSVC~\cite{yang2024adaptive}}      & \multicolumn{1}{c}{TOMM'24} & \multicolumn{1}{c}{RA} & 32  & \multicolumn{1}{c|}{96}& 50.0 &56.4 &26.8& 19.2 &37.3 &51.7 &40.3\\
  \rowcolor{lightgray}\multicolumn{1}{c}{DCVC-B~\cite{sheng2025bi}}      & \multicolumn{1}{c}{TMM'25} & \multicolumn{1}{c}{RA}& 32  & \multicolumn{1}{c|}{96} &19.1 &29.0 &-8.4 &-11.2 &18.4 &25.0 &12.0\\
  \multicolumn{1}{c}{L-LBVC~\cite{zhai2025llbvc}}      & \multicolumn{1}{c}{DCC'25} & \multicolumn{1}{c}{RA} & 32  & \multicolumn{1}{c|}{96}& 24.2&33.9 &-11.2 &-4.3& 10.1 &34.5 &14.5 \\
  \rowcolor{lightgray}\multicolumn{1}{c}{BiECVC}      & \multicolumn{1}{c}{Ours} & \multicolumn{1}{c}{RA}  & 32  & \multicolumn{1}{c|}{96}& \textcolor{purple}{$\bm{-6.8}$}   & \textcolor{purple}{$\bm{-3.9}$} & \textcolor{purple}{$\bm{-28.6}$}  & \textcolor{purple}{$\bm{-20.2}$} &    \textcolor{purple}{$\bm{-16.7}$} &  \textcolor{purple}{$\bm{-4.0}$}   & \textcolor{purple}{$\bm{-13.4}$}\\
  \midrule
  \multicolumn{1}{c}{B-CANF~\cite{chen2024bcanf}}      & \multicolumn{1}{c}{TCSVT'23} & \multicolumn{1}{c}{RA}  & 32 &  \multicolumn{1}{c|}{97}  & 73.1 &84.9 &43.3 &73.5& 72.8 &86.3 &72.3\\
  \rowcolor{lightgray}\multicolumn{1}{c}{DCVC-B~\cite{sheng2025bi}}      & \multicolumn{1}{c}{TMM'25} & \multicolumn{1}{c}{RA} & 32 &  \multicolumn{1}{c|}{97} &10.6 &19.3 &-13.0 &-13.0 &17.7 &23.3 &7.5\\
   \multicolumn{1}{c}{BiECVC}      & \multicolumn{1}{c}{Ours} & \multicolumn{1}{c}{RA} & 32 &  \multicolumn{1}{c|}{97} & \textcolor{purple}{$\bm{-14.8}$}   & \textcolor{purple}{$\bm{-9.0}$} & \textcolor{purple}{$\bm{-31.9}$}  & \textcolor{purple}{$\bm{-26.2}$} &    \textcolor{purple}{$\bm{-17.5}$} &  \textcolor{purple}{$\bm{-4.6}$}   & \textcolor{purple}{$\bm{-17.3}$}\\
   \midrule
   \rowcolor{lightgray}\multicolumn{1}{c}{VTM-13.2 LDB~\cite{bross2021overview}}     & \multicolumn{1}{c}{---}               & \multicolumn{1}{c}{LD}       & 64 &  \multicolumn{1}{c|}{64}       &26.0  &20.9  &22.1 & 9.1  &28.4  &30.2 & 22.8    \\  
   \multicolumn{1}{c}{DCVC-TCM~\cite{sheng2022temporal}}     & \multicolumn{1}{c}{TMM'22}              & \multicolumn{1}{c}{LD}         & 64 &  \multicolumn{1}{c|}{64}          &88.7 &119.1 &79.4 &151.1 &86.8 &82.1 &101.2 \\
   \rowcolor{lightgray}\multicolumn{1}{c}{DCVC-HEM~\cite{li2022hybrid}}    & \multicolumn{1}{c}{ACMMM'22}                   & \multicolumn{1}{c}{LD}       & 64 &  \multicolumn{1}{c|}{64}     &30.7 &49.1 &21.6 &56.5 &22.6 &27.5 &34.7      \\
   \multicolumn{1}{c}{DCVC-DC~\cite{li2023neural}}      & \multicolumn{1}{c}{CVPR'23}                   & \multicolumn{1}{c}{LD}      & 64 &  \multicolumn{1}{c|}{64}     &8.2 &15.3 &-7.2 &-2.4 &-0.5 &7.7 &3.5        \\
   \rowcolor{lightgray}\multicolumn{1}{c}{DCVC-FM~\cite{li2024neural}}        & \multicolumn{1}{c}{CVPR'24}                 & \multicolumn{1}{c}{LD}    & 64 &  \multicolumn{1}{c|}{64} &12.1 &11.9 &-8.4 &-13.2 &-0.3 &9.9 &2.0      \\
   \multicolumn{1}{c}{ECVC~\cite{jiang2025ecvc}}      & \multicolumn{1}{c}{CVPR'25} & \multicolumn{1}{c}{LD}  & 64 &  \multicolumn{1}{c|}{64}  &-7.4 &-1.5 &-22.2 &-3.8& -18.1 &-5.1 &-9.7\\
   \rowcolor{lightgray}\multicolumn{1}{c}{B-CANF~\cite{chen2024bcanf}}      & \multicolumn{1}{c}{TCSVT'23} & \multicolumn{1}{c}{RA}  & 64 &  \multicolumn{1}{c|}{64} & 94.2 &106.3 &64.9 &91.0 &89.5 &102.5 &91.4\\
   \multicolumn{1}{c}{DCVC-B~\cite{sheng2025bi}}      & \multicolumn{1}{c}{TMM'25} & \multicolumn{1}{c}{RA} & 64 &  \multicolumn{1}{c|}{64}&22.3 &25.3 &-9.3 &-23.4 &29.8 &37.5 &13.7\\
   \rowcolor{lightgray}\multicolumn{1}{c}{BiECVC}      & \multicolumn{1}{c}{Ours} & \multicolumn{1}{c}{RA} & 64 &  \multicolumn{1}{c|}{64}&  \textcolor{purple}{$\bm{-8.3}$}   & \textcolor{purple}{$\bm{-6.8}$} & \textcolor{purple}{$\bm{-30.5}$}  & \textcolor{purple}{$\bm{-39.7}$} &    \textcolor{purple}{$\bm{-12.0}$} &  \textcolor{purple}{$\bm{-3.2}$}   & \textcolor{purple}{$\bm{-15.7}$}\\
   \midrule
   \multicolumn{1}{c}{B-CANF~\cite{chen2024bcanf}}      & \multicolumn{1}{c}{TCSVT'23}& \multicolumn{1}{c}{RA}  & 64 &  \multicolumn{1}{c|}{65}  &75.9 &87.3 &49.4 &71.4 &71.4 &92.2 &74.6\\
  \rowcolor{lightgray}\multicolumn{1}{c}{DCVC-B~\cite{sheng2025bi}}      & \multicolumn{1}{c}{TMM'25} & \multicolumn{1}{c}{RA}  & 64 &  \multicolumn{1}{c|}{65} &12.3 &13.9 &-14.6 &-29.4 &16.6 &26.8& 4.3\\
 \multicolumn{1}{c}{BiECVC}      & \multicolumn{1}{c}{Ours} & \multicolumn{1}{c}{RA}  & 64 &  \multicolumn{1}{c|}{65} &  \textcolor{purple}{$\bm{-14.3}$}   & \textcolor{purple}{$\bm{-13.0}$} & \textcolor{purple}{$\bm{-35.3}$}  & \textcolor{purple}{$\bm{-43.9}$} &    \textcolor{purple}{$\bm{-17.0}$} &  \textcolor{purple}{$\bm{-2.7}$}   & \textcolor{purple}{$\bm{-21.0}$}\\
   \bottomrule  
\end{tabular}}
\begin{tablenotes}    
  \small               
\item Several methods are only compared under the intra period of 32 with 96 frames, due to the unavailability of their source code and model weights.
\end{tablenotes}       
\end{threeparttable}}
\caption{{BD-Rate $(\%)$~\cite{bjontegaard2001calculation} comparison for PSNR (dB). The anchor is \textbf{VTM-13.2 RA}.}}
\label{tab:rd}   
\end{table*}
\section{Experiments}
\subsection{Experimental Setup}
\subsubsection{Training Details}
The proposed BiECVC is implemented using PyTorch 2.2.2~\cite{paszke2019pytorch} and trained on the Vimeo-90K training split~\cite{xue2019video} 
and the BVI-AOM dataset\footnote[2]{Serval sequences are removed to prevent potential data leakage.}~\cite{nawala2024aom} using four Tesla A100-80G GPUs.
Following ECVC~\cite{jiang2025ecvc}, we adopt the multi-stage training strategy with long-sequence fine-tuning.
During training, sequences are randomly cropped into $256\times 256$ patches with a batch size of 4. 
In the finetuning stage, we incorporate $33$-frame sequences and set the learning rate to $10^{-6}$.
The optimization objective is defined as $\mathcal{L} = \mathcal{R} + w_{k} \times \lambda \times \mathcal{D}$, 
where $\mathcal{R}$ represents the bit rate and $\mathcal{D}$ denotes distortion.
For variable bit-rate training, $\lambda$ is sampled from $\{85, 170, 380, 840\}$.
We use hierarchical quality control with weight factors $w_k = \{1.4, 1.4, 0.7, 0.5, 0.5\}$\cite{sheng2025bi}, where $k$ denotes the layer index.
For intra-frame coding, we adopt the intra-frame codec of DCVC-DC~\cite{li2024neural}.
\begin{figure}
  \includegraphics[width=0.49\textwidth]{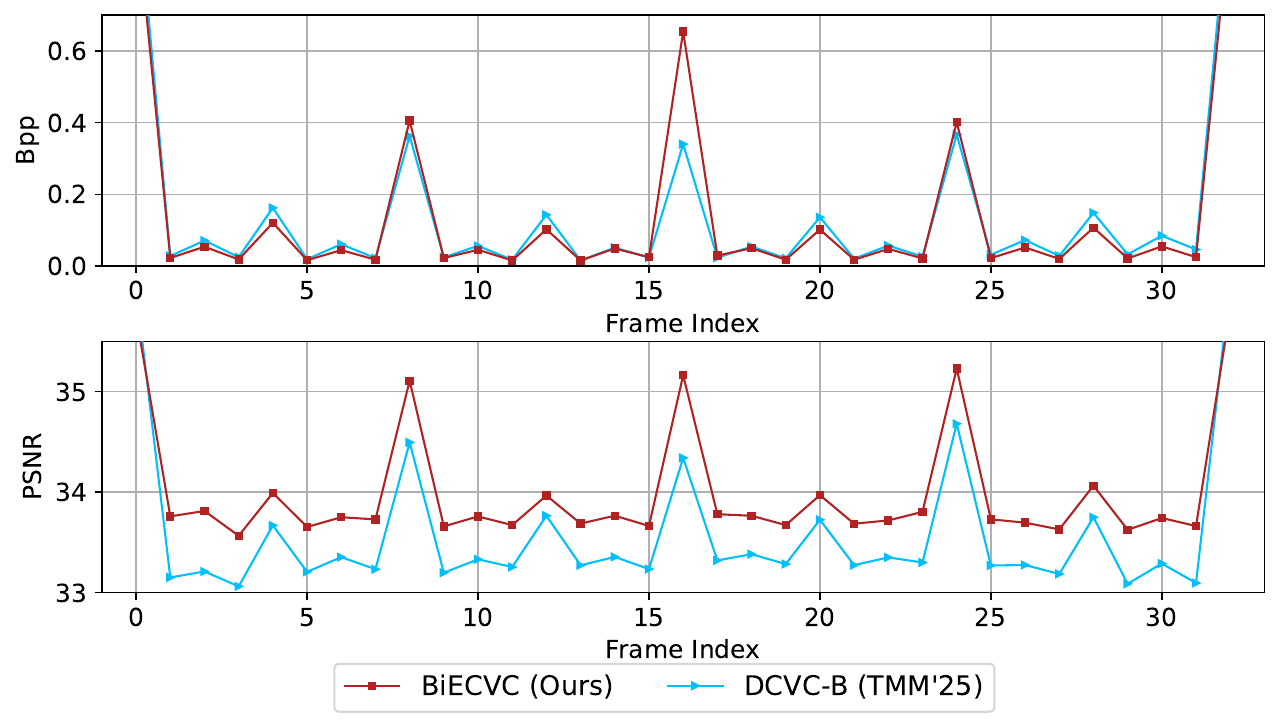}
  \caption{Rate Allocations on "BQTerrace" from HEVC B. The average bpps are around 0.09.}
  \label{fig:rate}
\end{figure}
\begin{figure}
  \includegraphics[width=0.38\textwidth]{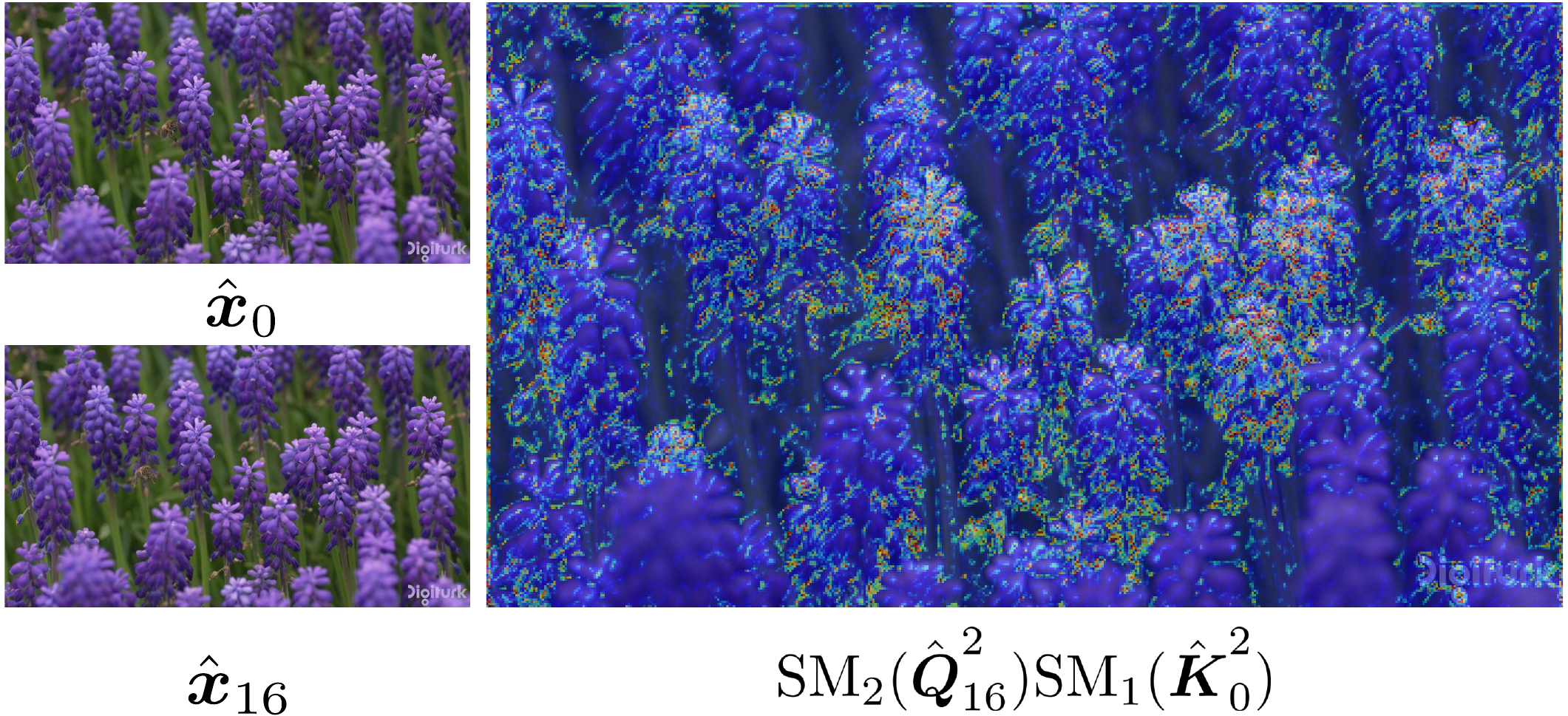}
  \caption{Visualization of attention maps with a query selected from flower regions.
  The frames are from "HoneyBee" sequence in UVG~\cite{mercat2020uvg} dataset.}
  \label{fig:attn}
\end{figure}
\subsubsection{Evaluation Details}
Following prior works~\cite{li2021deep,li2022hybrid,sheng2022temporal,li2024neural,li2023neural,jiang2024lvc,lu2024deep},
BiECVC is evaluated on HEVC datasets~\cite{bossen2013common}, including Class B, C, D, E, as well as UVG~\cite{mercat2020uvg} 
and MCL-JCV~\cite{wang2016mcl}.
To comprehensively assess its performance, we compare BiECVC with both low-delay and random access methods:
\begin{itemize}
  \item {Low-delay methods}: DCVC-TCM~\cite{sheng2022temporal}, 
  DCVC-HEM~\cite{li2022hybrid}, DCVC-DC~\cite{li2023neural}, DCVC-FM~\cite{li2024neural}, and ECVC~\cite{jiang2025ecvc}.
  \item {Random access methods}: B-CANF~\cite{chen2024bcanf}, UCVC~\cite{yang2024ucvc}, 
  APSVC~\cite{yang2024adaptive}, DCVC-B~\cite{sheng2025bi} and L-LBVC~\cite{zhai2025llbvc}.
\end{itemize}
Performance is measured using PSNR in RGB color format.
We evaluate BiECVC under IP32 with 96 frames and IP64 with 64 frames\footnote[3]{Since serval sequences of MCL-JCV only have 125 frames.} to ensure fair comparison with low delay methods.
Since RA relies on future intra frame, we duplicate the previous intra frame ($\hat{\boldsymbol{x}}_{64}$ or $\hat{\boldsymbol{x}}_{0}$) 
as a proxy for the unavailable future intra frame ($\hat{\boldsymbol{x}}_{96}$ or $\hat{\boldsymbol{x}}_{64}$), 
avoiding future information leakage.
To fully assess RA performance, we also report results on 97 and 65 frames, where actual future intra frames are accessible.
The Bjontegaard Delta-Rate (BD-Rate) metric~\cite{bjontegaard2001calculation} is used to rank the methods.
\begin{table}  
  \centering
  \small
  \setlength{\tabcolsep}{1.5mm}{
  \renewcommand\arraystretch{1}
  \begin{tabular}{@{}cccccccccccccc@{}}
  \toprule
  \multicolumn{1}{c}{\multirow{1}{*}{Method}}                        & Type    & \multicolumn{1}{c}{Params} & \multicolumn{1}{c}{kMACs/Pixel} & \multicolumn{1}{c}{ET} & \multicolumn{1}{c}{DT}\\ \midrule                                      
  \multicolumn{1}{c}{DCVC-HEM~\cite{li2022hybrid}}                     & LD                  & $50.9$       & $1767.19$ & $0.089$       & $0.136$ \\
  \multicolumn{1}{c}{DCVC-DC~\cite{li2023neural}}                     & LD                  & $50.8$       & $1332.96$ & $0.086$       & $0.116$ \\
  \multicolumn{1}{c}{DCVC-FM~\cite{li2024neural}}                     & LD                    & $44.9$       & $1125.95$ &  $0.085$    & $0.121$  \\
  \multicolumn{1}{c}{ECVC~\cite{jiang2025ecvc} }    & LD  & 61.9    & $1677.38$ &  $0.113$    & $0.149$ \\\midrule
  \multicolumn{1}{c}{B-CANF~\cite{sheng2025bi} }    & RA  & 46.7    & $2786.58$ &  $0.183$    & $0.177$ \\
  \multicolumn{1}{c}{DCVC-B~\cite{sheng2025bi} }    & RA  & 55.4    & $2920.31$ &  $0.191$    & $0.186$ \\
  \multicolumn{1}{c}{BiECVC w/o Cache }    & RA  & 63.9    & $2850.00$ &  $0.218$    & $0.207$ \\
  \multicolumn{1}{c}{BiECVC }    & RA  & 63.9    & $2671.20$ &  $0.179$    & $0.171$ \\
  \bottomrule  
\end{tabular}}
\caption{{Complexity comparison on 720p sequence.
"Params"(M) is the number of parameters. "kMACs/Pixel" is evaluated during forward. 
"ET"(s) and "DT"(s) refer to the average encoding and decoding time per frame.}}
\label{tab:complex} 
\end{table}
\begin{table*}
  \setlength{\tabcolsep}{3.5mm}{
  \renewcommand\arraystretch{1}
  \centering
  \small
  \begin{tabular}[t]{l|ccccccc}
  \toprule
  Variant &  {HEVC B}  & HEVC C & HEVC D & HEVC E & UVG & MCL-JCV & Average\\
  \midrule
  Base     & 14.0 &16.5 &-15.7 &0.6 &7.8 &12.8 &6.0\\
  MSE + BLCD       & 9.3 & 13.2  & -19.4 & -5.7 & 2.7 & 8.4 & 1.4\\
  MSE + BLCD + BNLCD    & -1.2  & 1.2 & -24.3 & -18.3  & -11.9 & 0.7 & -9.0\\
  MSE + BLCD + BNLCD + BCG    &  -6.8 &-3.9 &-28.6 &-20.2 &-16.7 &-4.0 &-13.4   \\
  \bottomrule
  \end{tabular}}
  \caption{Ablation Studies on HEVC B$\sim$E, UVG and MCL-JCV. The metric is BD-Rate (\%) w.r.t VTM-13.2 RA.}
  \label{tab:ablation}
  \end{table*}
\subsection{Comparisons with Previous SOTA Methods}
\subsubsection{Rate-Distortion Performance}
The performances\footnote[4]{The MS-SSIM results, as well as the results on all sequences under the IP64 setting with 192/193 frames (excluding MCL-JCV), are presented in the supplementary material.} are presented in Figure~\ref{fig:rd} and Table~\ref{tab:rd}.
In Figure~\ref{fig:rd}, 16 data points of BiECVC are plotted to demonstrate the ability of variable rate.\par
\textbf{Under IP32}. 
We evaluate BiECVC with 96 and 97 frames.
On 96 frames, BiECVC achieves significant improvements over SOTA non-learned codecs, 
including VTM-13.2 under both LDB and RA configurations.
On average, BiECVC outperforms VTM-13.2 RA by 13.4\% in BD-rate.
Compared to low-delay learned methods, BiECVC consistently surpasses ECVC~\cite{jiang2025ecvc} across all datasets, with an average gain of 17\%.
Furthermore, BiECVC outperforms the SOTA random access method DCVC-B~\cite{sheng2025bi} by a large margin of 25.4\% on average.
On 97 frames, we compare BiECVC only against existing BVCs and VTM-13.2 RA.
In this setting, BiECVC achieves even greater improvements, reducing BD-rate by 17.3\% on average relative to VTM-13.2 RA.
\par
\textbf{Under IP64}.
On 64 frames, BiECVC outperforms VTM-13.2 RA by 15.7\% in BD-rate.
On 65 frames, BiECVC achieves even greater improvements, reducing BD-rate by 21\% on average relative to VTM-13.2 RA.
Under the IP64 setting, larger motion in lower-layer frames leads to less informative reference contexts for bidirectional prediction, 
limiting performance gains. This also occurs in traditional codecs, where the advantage of VTM RA over VTM LD diminishes under longer IPs. 
It explains why BiECVC underperforms ECVC on certain sequences.
Nonetheless, BiECVC still achieves an average bit-rate reduction of 6\% compared to ECVC.\par
To our knowledge, this is the \textit{first} learned video compression model to surpass VTM-13.2 RA in rate-distortion performance.
Notably, BiECVC consistently outperforms VTM-13.2 RA across all datasets, demonstrating its strong generalization ability.
\begin{figure}
  \includegraphics[width=0.485\textwidth]{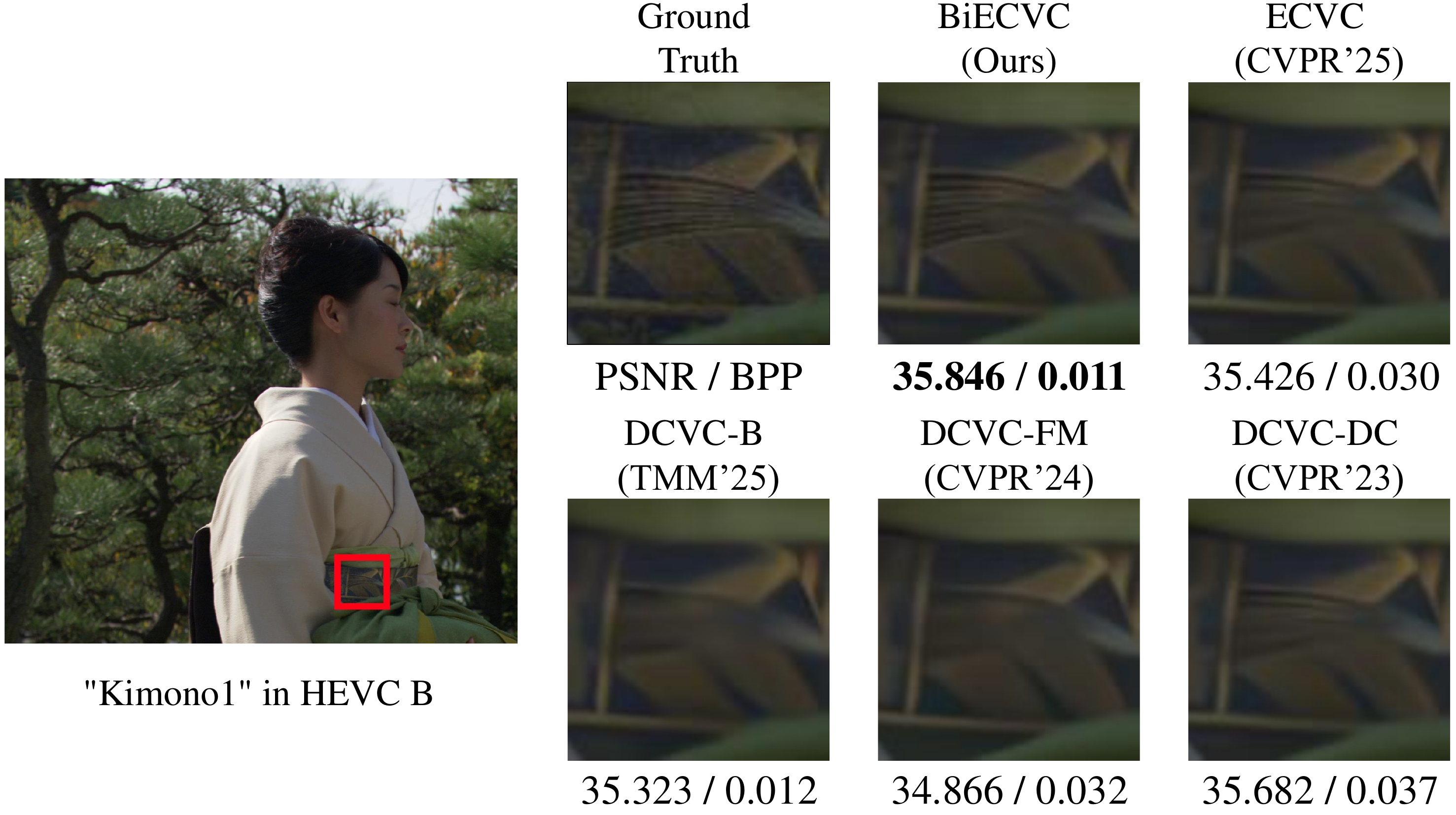}
  \caption{Visual quality comparison on reconstruction frames and ground truth.}
  \label{fig:subjective}
\end{figure}
\subsubsection{Rate Allocation Analysis}
Since BiECVC is designed for B-frame coding, we compare it with DCVC-B~\cite{sheng2025bi} in terms of rate allocation. 
The results are shown in Figure~\ref{fig:rate}.
At lower layers (\textit{e.g.}, frame 16), BiECVC allocates more bits, whereas at higher layers, it allocates fewer bits compared to DCVC-B. 
This strategy is more reasonable because lower-layer 
frames serve as references for higher-layer frames, making it beneficial to allocate more bits to improve their quality.
\subsubsection{Visual Quality Comparison}
The visual quality comparison is shown in Figure~\ref{fig:subjective}.
Compared with ECVC, BiECVC achieves a 0.4 dB improvement in PSNR while consuming fewer bits.
In terms of visual quality, our BiECVC demonstrates noticeable improvements over prior state-of-the-art methods~\cite{jiang2025ecvc,li2024neural,li2023neural,sheng2025bi}.
\subsubsection{Complexity Analysis}
BiECVC is compared with both low-delay~\cite{li2023neural,li2024neural,jiang2025ecvc,li2022hybrid} and 
random access~\cite{chen2024bcanf,sheng2025bi} baselines in terms of complexity, as shown in Table~\ref{tab:complex}. 
While bidirectional prediction inherently involves higher complexity due to multiple motion estimations and compensations, 
BiECVC achieves significantly better performance than low-delay methods~\cite{li2023neural,li2024neural,jiang2025ecvc,li2022hybrid} despite its higher complexity. Moreover, 
compared with other bidirectional approaches such as B-CANF~\cite{chen2024bcanf} and DCVC-B~\cite{sheng2022temporal}, 
BiECVC achieves lower complexity thanks to the proposed feature cache.
\subsection{Ablation Studies}
In the ablation studies, all models are optimized using the same setting, and PSNR is adopted to evaluate distortion.
The results are summarized in Table~\ref{tab:ablation}.
\textit{The baseline model ("Base") does not incorporate bidirectional context diversification or weighting; instead, 
it directly warps two reference frames for conditional coding.}
According to Table~\ref{tab:ablation}, introducing BLCD improves performance.
BNLCD contributes the most to the overall performance, likely due to its non-local receptive field, which better handles various types of motion.
To further analyze the non-local context, we visualize the attention map in Figure~\ref{fig:attn}.
When a query from the flower region is selected, the attention scores are high across multiple flower areas,
indicating that non-local dependencies are effectively captured.
BCG provides a relatively smaller performance gain on HEVC Class E, possibly because the sequences in HEVC E contain only small movements.
The ablation studies demonstrate the effectiveness of our proposed techniques for BiECVC. 
\section{Conclusion}
In this paper, we present BiECVC, a bidirectional video compression framework that leverages local and non-local context diversification along with context gating.
To enhance local context diversity, we incorporate high-quality features from lower layers. These features are aligned using existing decoded motion vectors, eliminating the need for additional motion estimation and compression. 
For non-local context diversification, we adopt a linear attention mechanism to strike a balance between performance and computational efficiency.
Furthermore, we introduce a bidirectional context gating mechanism that dynamically emphasizes beneficial information while suppressing harmful noise, drawing inspiration from recent advancements in linear-time autoregressive language modeling~\cite{yang2024gated,qin2023hierarchically}.
To accelerate inference, we integrate a feature cache that reuses precomputed features.
Experimental results demonstrate that BiECVC achieves SOTA performance. To the best of our knowledge, this is the \textit{first} learned video codec to surpass VTM 13.2 RA~\cite{bross2021overview} across all datasets.
However, since BiECVC builds upon the DCVC-DC~\cite{li2023neural} and ECVC~\cite{jiang2025ecvc} frameworks, its speed remains far from real-time. In future work, we plan to integrate the proposed techniques into DCVC-RT~\cite{jia2025practical} to enable practical bidirectional video compression.
\balance
\bibliographystyle{ACM-Reference-Format}
\bibliography{main}
\appendix
\onecolumn
\section{Test Settings}
We evaluate both learned video codecs (LVCs) and traditional codecs in the RGB color space.
Frames encoded in YUV are converted to RGB using the BT.601 standard.
For obtaining the rate-distortion (RD) data of VTM-13.2 under LDB and RA configurations~\cite{bross2021overview},
the following commands are used:\par
\quad\par
\begin{minipage}[t]{0.48\textwidth}
\centering
\begin{verbatim}
# VTM-13.2 LDB
./EncoderAppStatic
-c encoder_lowdelay_vtm.cfg
--InputFile={input file name}
--BitstreamFile={bitstream file name}
--DecodingRefreshType=2
--InputBitDepth=8 
--OutputBitDepth=8 
--OutputBitDepth=8
--InputChromaFormat=444
--FrameRate={frame rate} 
--FramesToBeEncoded={frame number}
--SourceWidth={width} 
--SourceHeight={height} 
--IntraPeriod={IP} 
--QP={qp} 
--Level=6.2
\end{verbatim}
\end{minipage}
\hfill
\begin{minipage}[t]{0.48\textwidth}
\centering
\begin{verbatim}
# VTM-13.2 RA
./EncoderAppStatic
-c encoder_randomaccess_vtm.cfg
--InputFile={input file name}
--BitstreamFile={bitstream file name}
--DecodingRefreshType=1
--InputBitDepth=8 
--OutputBitDepth=8 
--OutputBitDepth=8
--InputChromaFormat=444
--FrameRate={frame rate} 
--FramesToBeEncoded={frame number}
--SourceWidth={width} 
--SourceHeight={height} 
--IntraPeriod={IP} 
--QP={qp} 
--Level=6.2
\end{verbatim}
\end{minipage}
\section{Architectures}
\begin{figure*}
  \includegraphics[width=0.8\textwidth]{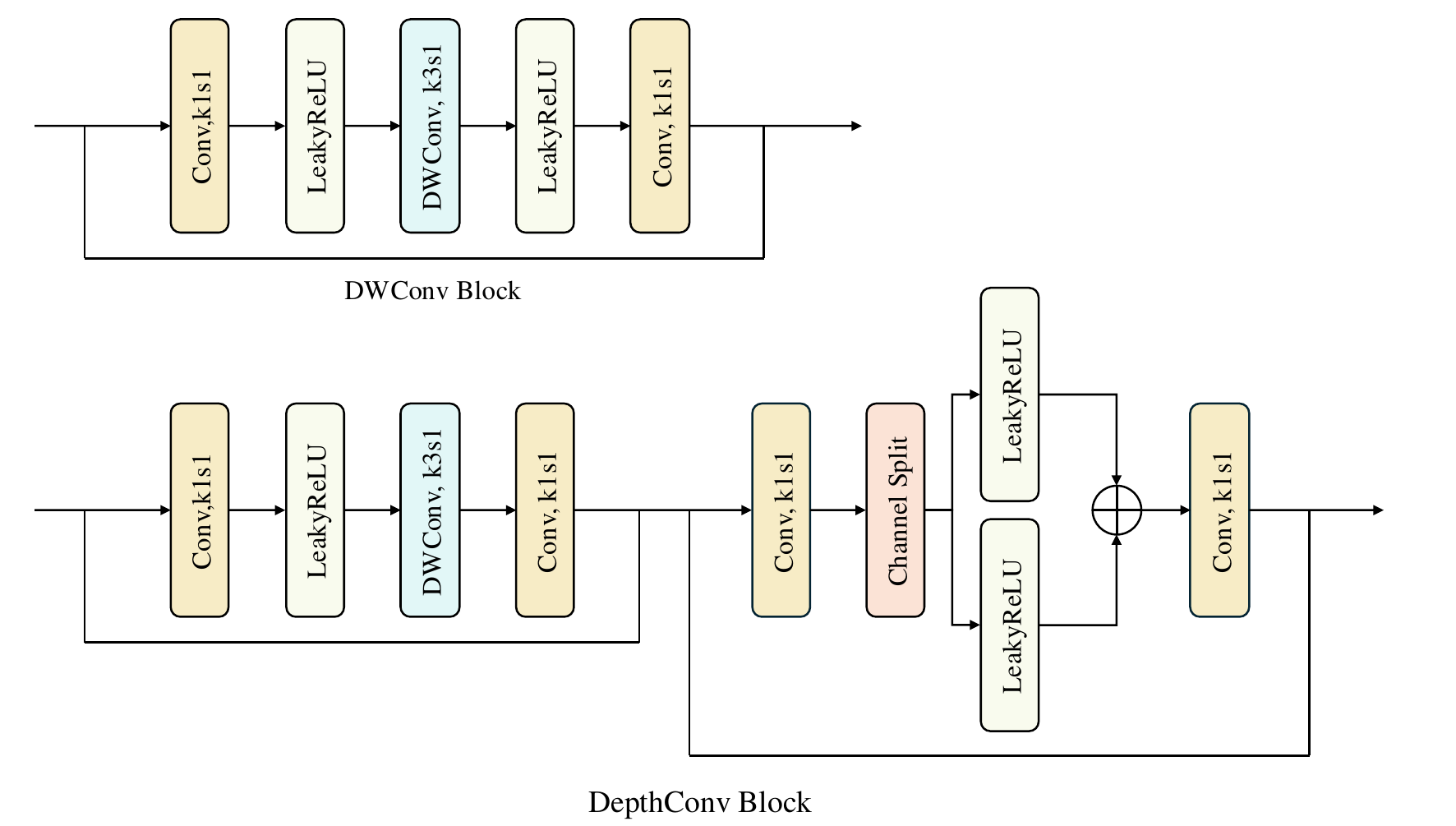}
  \caption{Architectures of DWConv Block~\cite{jiang2024llic} and DepthConv Block~\cite{li2024neural}.}
  \label{fig:dwc}
\end{figure*}

In this section, we describe the architectures of DWConv Block and Bidirectional Context Gating Module.
The architectures are presented in Figure~\ref{fig:dwc}.\par
The DWConv Block~\cite{jiang2024llic} first applies a $1\times 1$ convolution followed by a LeakyReLU activation to adjust channel dimensions. 
Then, a depthwise $3\times 3$ convolution captures spatial information efficiently, followed by another activation. 
Finally, a second $1\times 1$ convolution refines the output. A skip connection is added, which is either an identity mapping 
or a $1\times 1$ convolution if the input and output channel dimensions differ.\par
In Bidirectional Context Gating Module, the DepthConv Block~\cite{li2024neural} is employed for conditional coding and dimension reduction.
The DepthConv Block first applies a depthwise convolution layer (DepthConv) to extract spatial features with low computational cost, 
followed by a channel-wise feed-forward network (ConvFFN) to enhance channel interactions.\par
\section{Rate-Distortion Results}
\begin{figure*}
  \includegraphics[width=\textwidth]{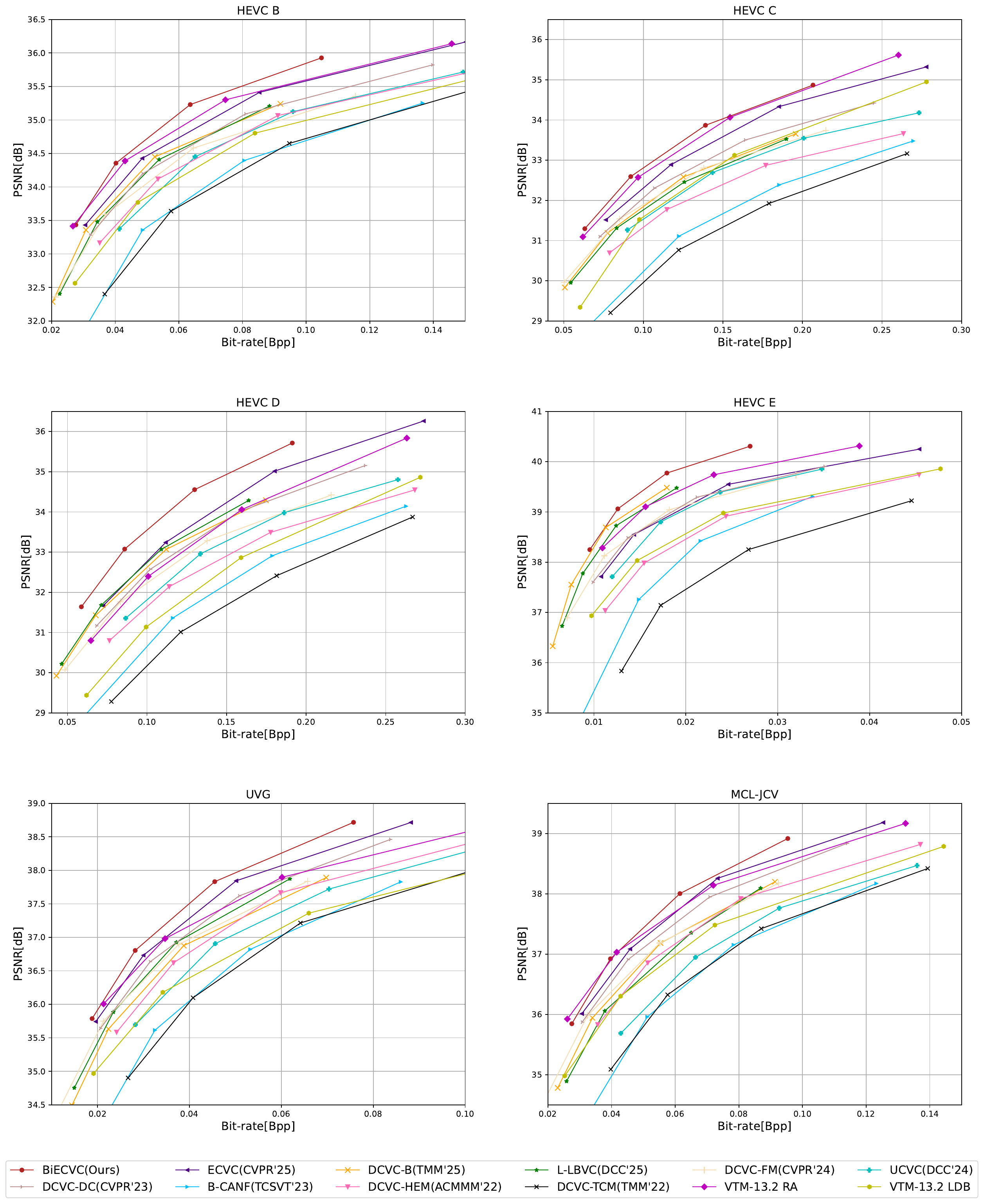}
  \caption{Rate-PSNR performances under intra period 32 with 96 frames.}
  \label{fig:ip32_96}
\end{figure*}

\begin{figure*}
  \includegraphics[width=\textwidth]{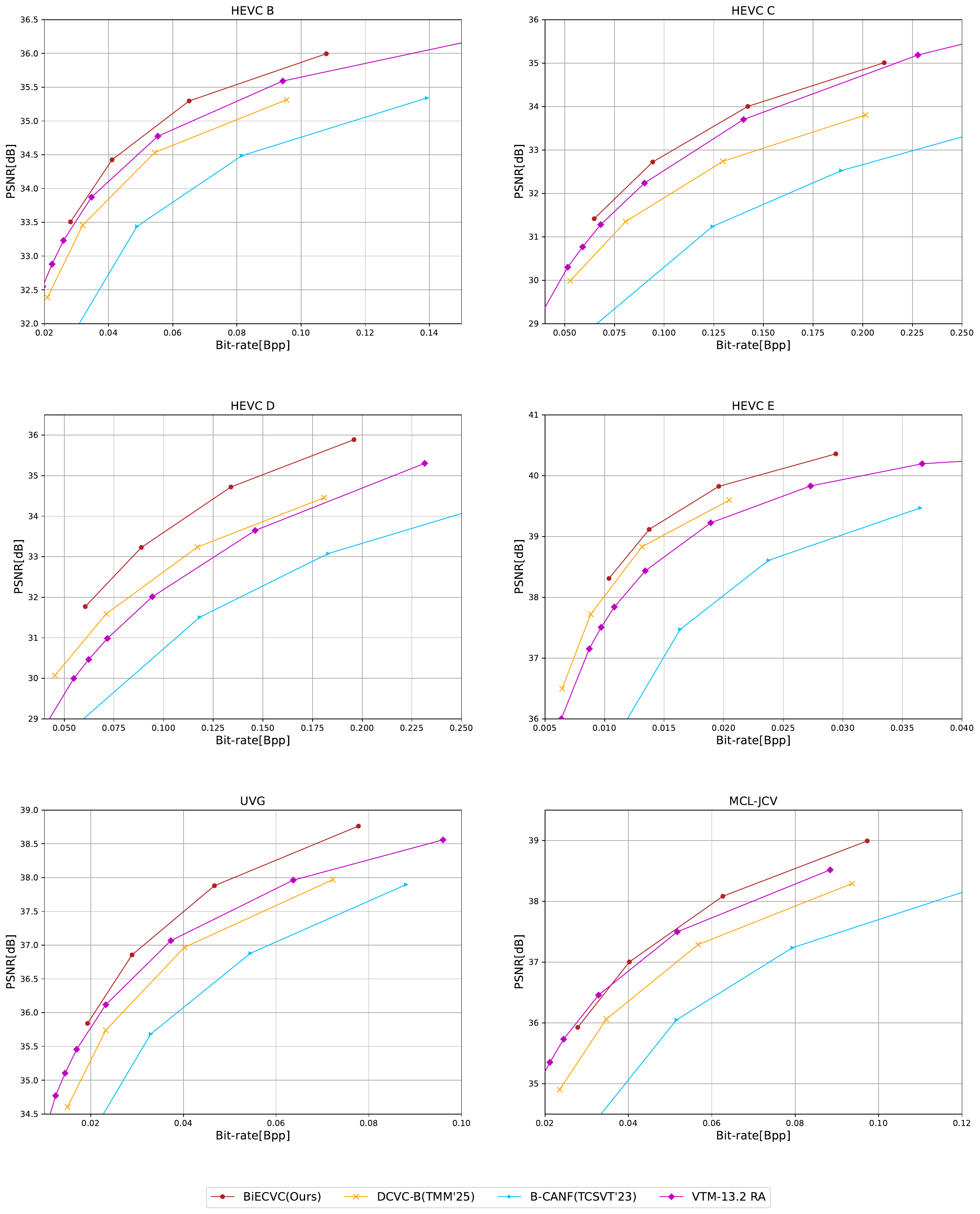}
  \caption{Rate-PSNR performances under intra period 32 with 97 frames.}
  \label{fig:ip32_97}
\end{figure*}

\begin{figure*}
  \includegraphics[width=\textwidth]{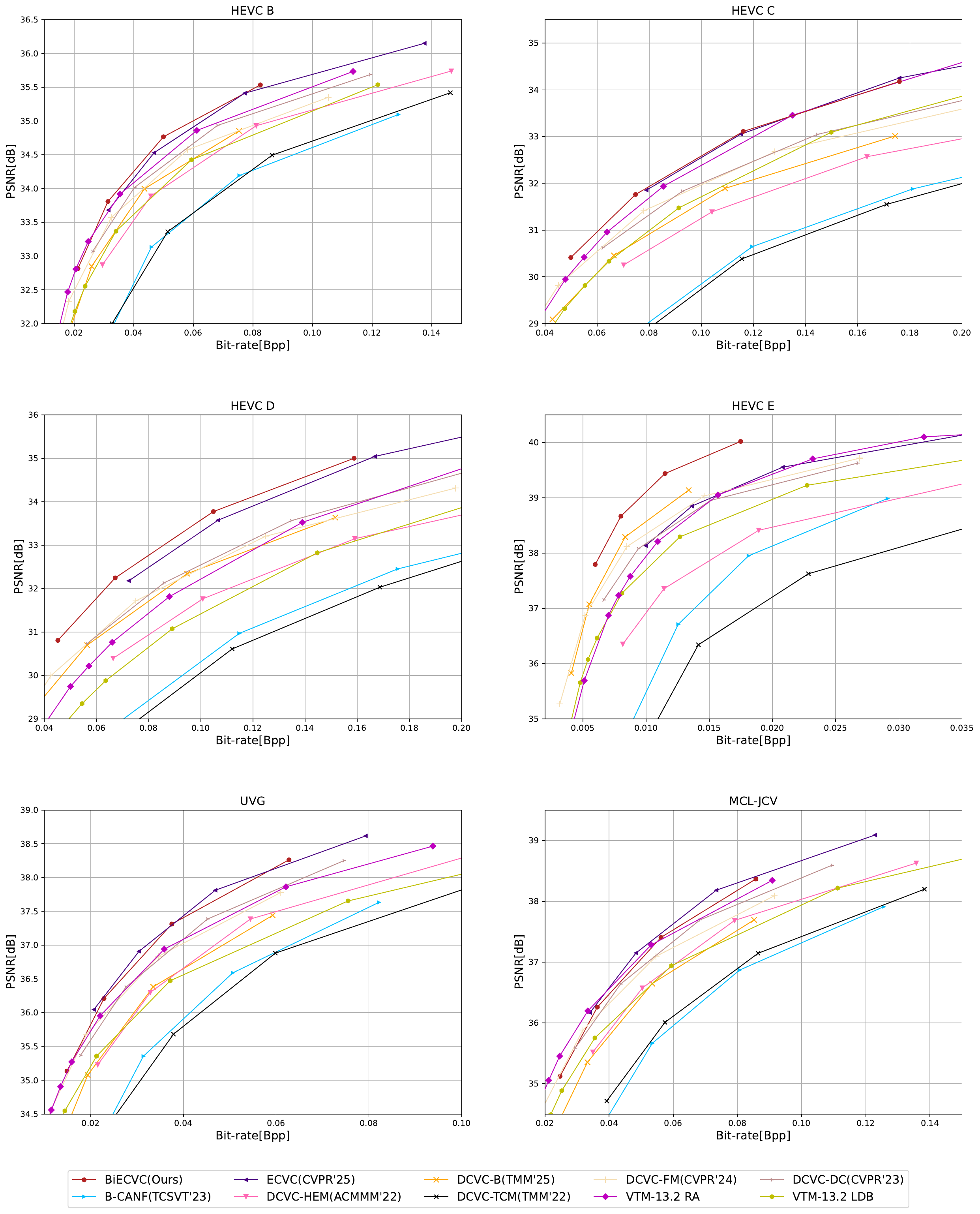}
  \caption{Rate-PSNR performances under intra period 64 with 64 frames.}
  \label{fig:ip64_64}
\end{figure*}

\begin{figure*}
  \includegraphics[width=\textwidth]{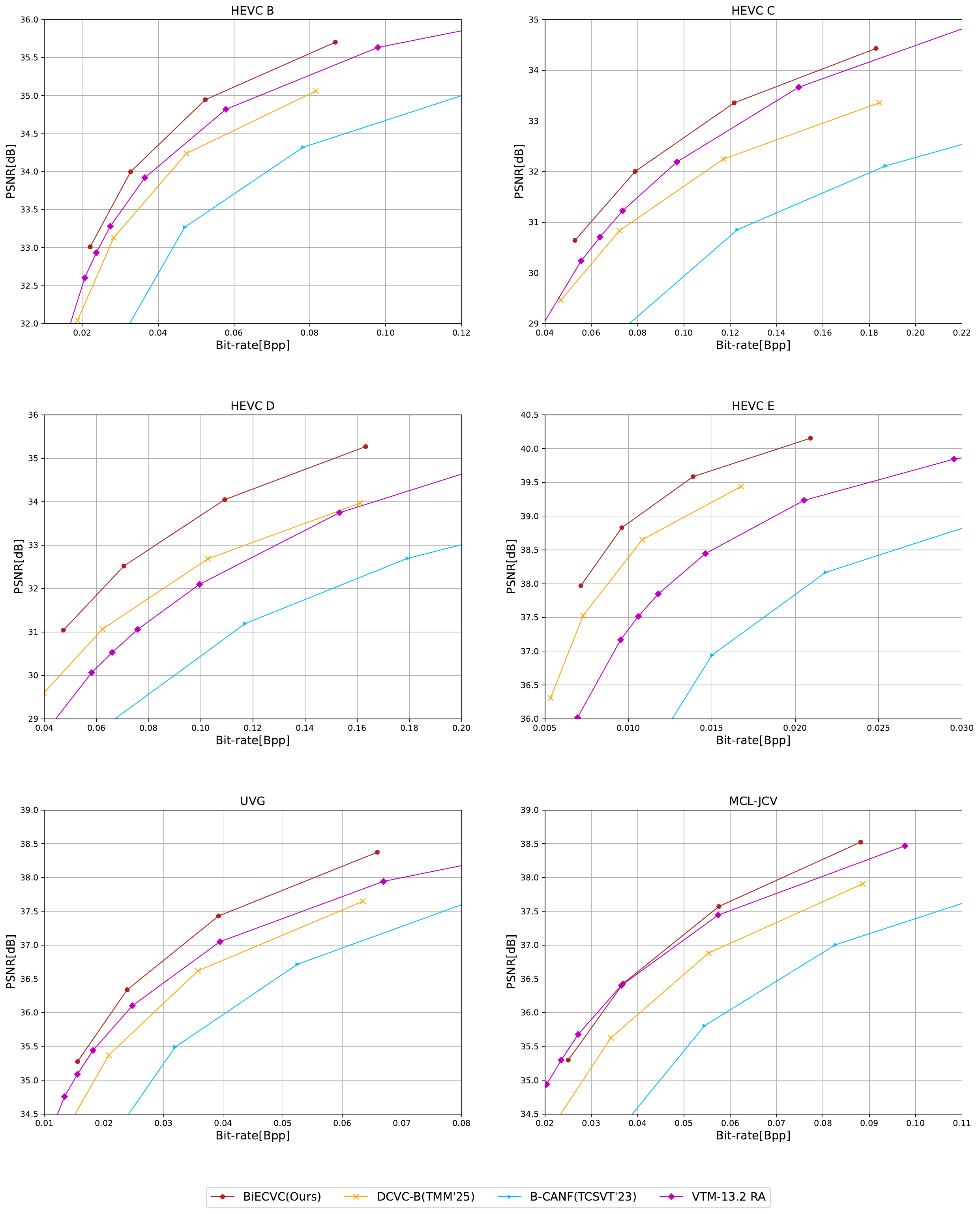}
  \caption{Rate-PSNR performances under intra period 64 with 65 frames.}
  \label{fig:ip64_65}
\end{figure*}

\begin{figure*}
  \includegraphics[width=\textwidth]{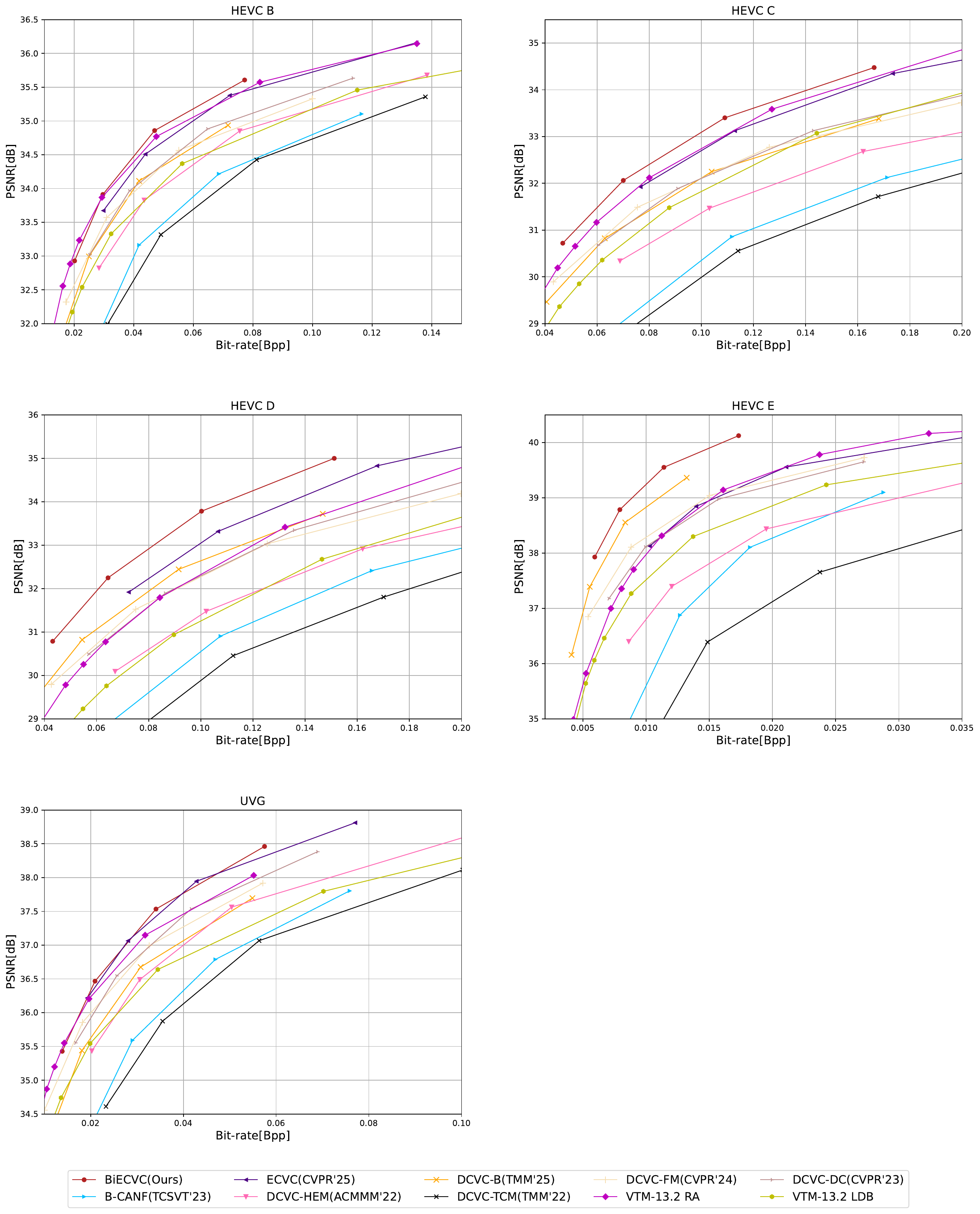}
  \caption{Rate-PSNR performances under intra period 64 with 192 frames.}
  \label{fig:ip64_192}
\end{figure*}

\begin{figure*}
  \includegraphics[width=\textwidth]{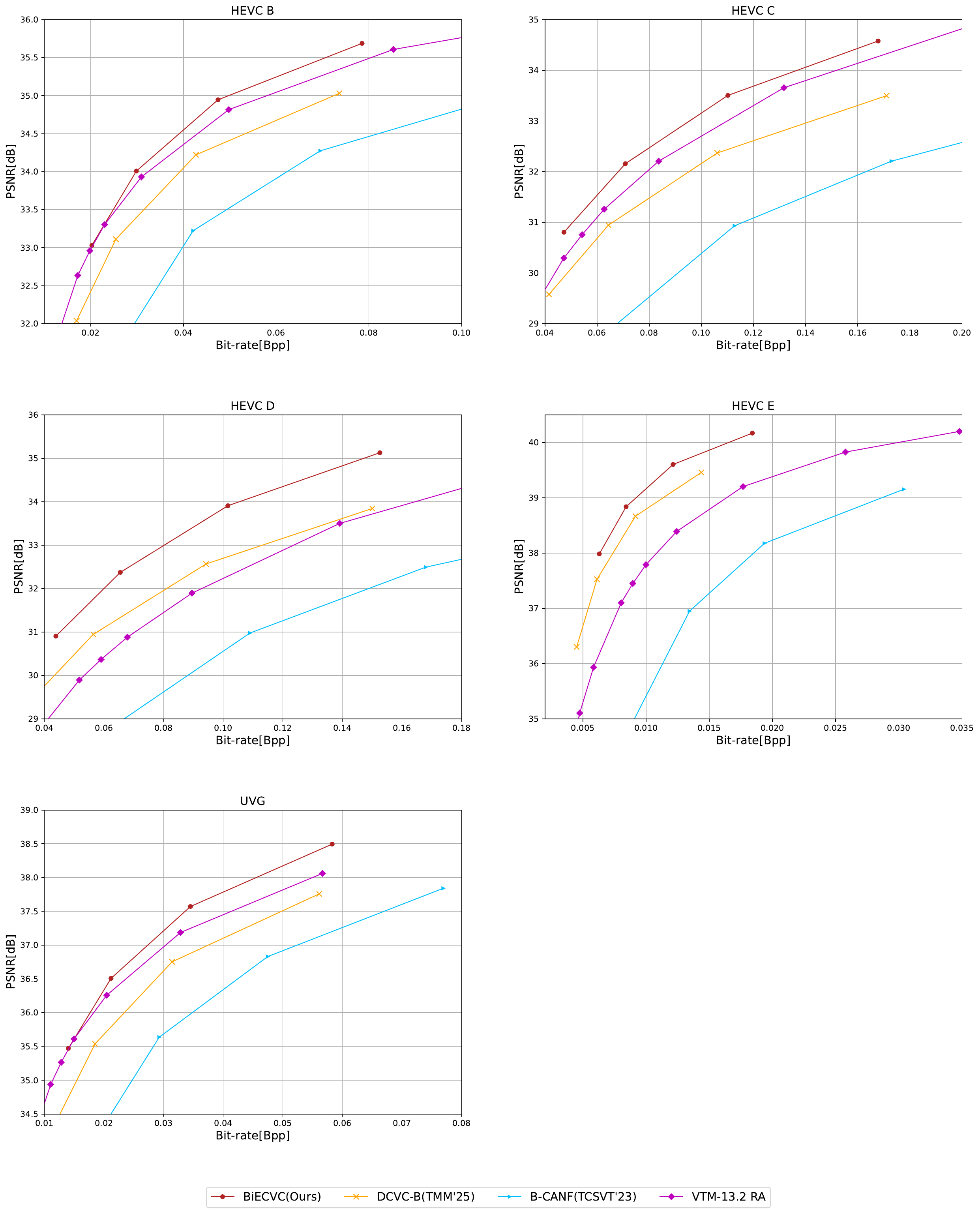}
  \caption{Rate-PSNR performances under intra period 64 with 193 frames.}
  \label{fig:ip64_193}
\end{figure*}

\begin{figure*}
  \includegraphics[width=\textwidth]{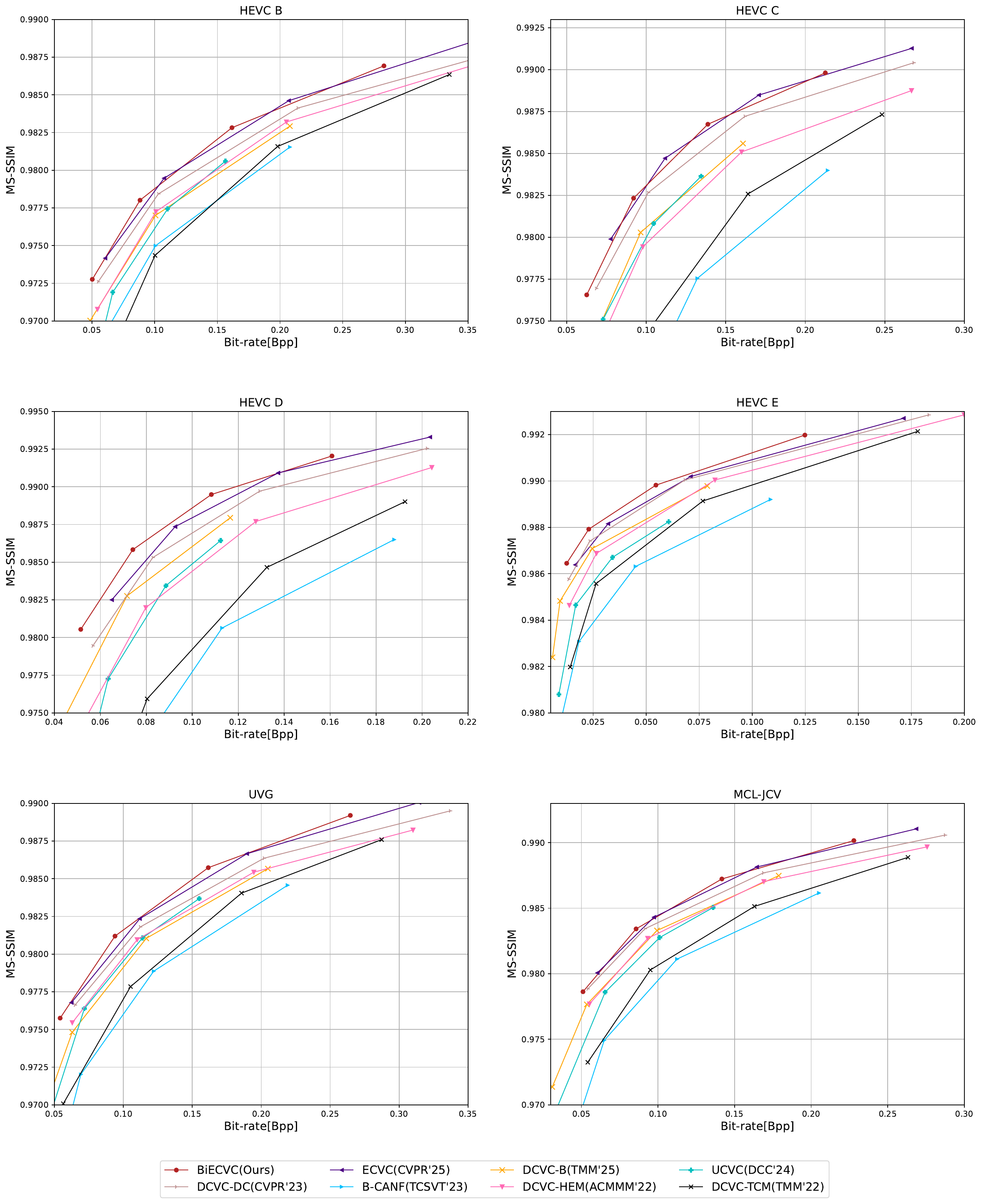}
  \caption{Rate-MS-SSIM~\cite{wang2003multiscale} performances under intra period 32 with 96 frames.}
  \label{fig:ip32_96_ssim}
\end{figure*}

For all baselines, we evaluate their performances using official code and weights.
To comprehensively assess its performance, we compare BiECVC with both low-delay and random access methods:
\begin{itemize}
  \item {Low-delay methods}: DCVC-TCM~\cite{sheng2022temporal}, 
  DCVC-HEM~\cite{li2022hybrid}, DCVC-DC~\cite{li2023neural}, DCVC-FM~\cite{li2024neural}, ECVC~\cite{jiang2025ecvc}, and VTM-13.2 LDB.
  \item {Random access methods}: B-CANF~\cite{chen2024bcanf}, UCVC~\cite{yang2024ucvc}, 
  DCVC-B~\cite{sheng2025bi}, L-LBVC~\cite{zhai2025llbvc} and VTM-13.2 RA.
\end{itemize}
The rate-distortion results are presented in Table~\ref{tab:rd_64_192},~\ref{tab:rd_ssim} and Figure~\ref{fig:ip32_96},~\ref{fig:ip32_97},~\ref{fig:ip64_64},~\ref{fig:ip64_65},~\ref{fig:ip64_192},~\ref{fig:ip64_193},~\ref{fig:ip32_96_ssim}.
\par
For MS-SSIM~\cite{wang2003multiscale} optimized BiECVC, we finetune the MSE-optimized model and sample $\lambda$ from $\{7.68, 15.36, 30.72, 61.44\}$ for variable rate.
\begin{table*}  
  \centering
  \setlength{\tabcolsep}{2.8mm}{
  \small
  \renewcommand\arraystretch{1.1}
    \begin{threeparttable}{
  \begin{tabular}{@{}cccccccccccccc@{}}
  \toprule
  \multicolumn{1}{c}{\multirow{2}{*}{Method}} & \multicolumn{1}{c}{\multirow{2}{*}{{Venue}}}              & \multicolumn{1}{c}{\multirow{2}{*}{{Type}}}    & \multicolumn{1}{c}{\multirow{1}{*}{{Intra}}} & \multicolumn{1}{c|}{\multirow{1}{*}{{Frame}}}           & \multicolumn{6}{c}{{BD-Rate (\%) w.r.t. VTM-13.2 RA~\cite{bross2021overview}}}  \\
  \multicolumn{1}{c}{}      & \multicolumn{1}{c}{}                           & \multicolumn{1}{c}{}        & \multicolumn{1}{c}{Period}        & \multicolumn{1}{c|}{Number}                   & \multicolumn{1}{c}{HEVC B}    & \multicolumn{1}{c}{HEVC C}  & \multicolumn{1}{c}{HEVC D} & \multicolumn{1}{c}{HEVC E}  & \multicolumn{1}{c}{UVG}& \multicolumn{1}{c}{Average} \\       \midrule                                      
   \multicolumn{1}{c}{VTM-13.2 LDB~\cite{bross2021overview}}     & \multicolumn{1}{c}{---}               & \multicolumn{1}{c}{LD}       & 64 &  \multicolumn{1}{c|}{192}       &42.8 &29.2 &31.8 &19.7 &39.3&32.6  \\  
   \rowcolor{lightgray}\multicolumn{1}{c}{DCVC-TCM~\cite{sheng2022temporal}}     & \multicolumn{1}{c}{TMM'22}              & \multicolumn{1}{c}{LD}         & 64 &  \multicolumn{1}{c|}{192}   &115.5& 132.6& 97.3& 160.7 &102.9 &121.8  \\
   \multicolumn{1}{c}{DCVC-HEM~\cite{li2022hybrid}}    & \multicolumn{1}{c}{ACMMM'22}                   & \multicolumn{1}{c}{LD}       & 64 &  \multicolumn{1}{c|}{192}     &51.7 &61.6 &36.6 &63.0 &32.8 &49.1   \\
   \rowcolor{lightgray}\multicolumn{1}{c}{DCVC-DC~\cite{li2023neural}}      & \multicolumn{1}{c}{CVPR'23}                   & \multicolumn{1}{c}{LD}      & 64 &  \multicolumn{1}{c|}{192}     &26.4 &24.5 &2.0 &3.1 &8.3 &12.9       \\
   \multicolumn{1}{c}{DCVC-FM~\cite{li2024neural}}        & \multicolumn{1}{c}{CVPR'24}                 & \multicolumn{1}{c}{LD}    & 64 &  \multicolumn{1}{c|}{192} &25.8 &18.6& -0.4 &-8.4 &10.6&9.2      \\
   \rowcolor{lightgray}\multicolumn{1}{c}{ECVC~\cite{jiang2025ecvc}}      & \multicolumn{1}{c}{CVPR'25} & \multicolumn{1}{c}{LD}  & 64 &  \multicolumn{1}{c|}{192}  &3.8 &3.8& -15.9 &1.0 &-9.9 &-3.4\\
   \multicolumn{1}{c}{B-CANF~\cite{chen2024bcanf}}      & \multicolumn{1}{c}{TCSVT'23} & \multicolumn{1}{c}{RA}  & 64 &  \multicolumn{1}{c|}{192} & 102.3 &105.8 &63.7 &85.7 &98.5& 91.2\\
  \rowcolor{lightgray}\multicolumn{1}{c}{DCVC-B~\cite{sheng2025bi}}      & \multicolumn{1}{c}{TMM'25} & \multicolumn{1}{c}{RA} & 64 &  \multicolumn{1}{c|}{192}&28.9 &20.3 &-12.2 &-29.6 &29.9 &7.5\\
   \multicolumn{1}{c}{BiECVC}      & \multicolumn{1}{c}{Ours} & \multicolumn{1}{c}{RA} & 64 &  \multicolumn{1}{c|}{192}&  \textcolor{purple}{$\bm{-3.1}$}   & \textcolor{purple}{$\bm{-10.7}$} & \textcolor{purple}{$\bm{-32.0}$}  & \textcolor{purple}{$\bm{-42.6}$} &    \textcolor{purple}{$\bm{-8.8}$} &  \textcolor{purple}{$\bm{-19.4}$}   \\
   \midrule
   \rowcolor{lightgray}\multicolumn{1}{c}{B-CANF~\cite{chen2024bcanf}}      & \multicolumn{1}{c}{TCSVT'23}& \multicolumn{1}{c}{RA}  & 64 &  \multicolumn{1}{c|}{193} & 93.4&  99.0&  56.8&  78.3&  92.3&  84.0 \\
 \multicolumn{1}{c}{DCVC-B~\cite{sheng2025bi}}      & \multicolumn{1}{c}{TMM'25} & \multicolumn{1}{c}{RA}  & 64 &  \multicolumn{1}{c|}{193} &21.7 &16.4 &-15.3 &-31.2 &25.1 &3.3\\
 \rowcolor{lightgray}\multicolumn{1}{c}{BiECVC}      & \multicolumn{1}{c}{Ours} & \multicolumn{1}{c}{RA}  & 64 &  \multicolumn{1}{c|}{193} &  \textcolor{purple}{$\bm{-8.9}$}   & \textcolor{purple}{$\bm{-13.7}$} & \textcolor{purple}{$\bm{-35.2}$}  & \textcolor{purple}{$\bm{-44.2}$} &    \textcolor{purple}{$\bm{-10.7}$} &  \textcolor{purple}{$\bm{-22.5}$}  \\
   \bottomrule  
\end{tabular}}   
\end{threeparttable}}
\caption{{BD-Rate $(\%)$~\cite{bjontegaard2001calculation} comparison for PSNR (dB). The anchor is \textbf{VTM-13.2 RA}.}}
\label{tab:rd_64_192}   
\end{table*}
\begin{table*}  
  \centering
  \setlength{\tabcolsep}{3.1mm}{
  \small
  \renewcommand\arraystretch{1.1}
    \begin{threeparttable}{
  \begin{tabular}{@{}cccccccccccccc@{}}
  \toprule
  \multicolumn{1}{c}{\multirow{2}{*}{Method}} & \multicolumn{1}{c}{\multirow{2}{*}{{Venue}}}              & \multicolumn{1}{c|}{\multirow{2}{*}{{Type}}}              & \multicolumn{7}{c}{{BD-Rate (\%) w.r.t. DCVC-TCM~\cite{sheng2022temporal}}}  \\
  \multicolumn{1}{c}{}      & \multicolumn{1}{c}{}                           & \multicolumn{1}{c|}{}                        & \multicolumn{1}{c}{HEVC B}    & \multicolumn{1}{c}{HEVC C}  & \multicolumn{1}{c}{HEVC D} & \multicolumn{1}{c}{HEVC E}  & \multicolumn{1}{c}{UVG~\cite{mercat2020uvg}} & \multicolumn{1}{c}{MCL-JCV~\cite{wang2016mcl}} & \multicolumn{1}{c}{Average} \\       \midrule                                      
  \multicolumn{1}{c}{DCVC-HEM~\cite{li2022hybrid}}    & \multicolumn{1}{c}{ACMMM'22}                   & \multicolumn{1}{c|}{LD}     &-19.3 &-24.8 &-27.1& -24.4 &-18.0 &-22.0 &-22.6             \\
  \rowcolor{lightgray}\multicolumn{1}{c}{DCVC-DC~\cite{li2023neural}}      & \multicolumn{1}{c}{CVPR'23}                   & \multicolumn{1}{c|}{LD}       &-26.6 &-38.2 &-40.1 &-38.8 &-23.8 &-30.1 &-32.9            \\
  \multicolumn{1}{c}{ECVC~\cite{jiang2025ecvc}}      & \multicolumn{1}{c}{CVPR'25} & \multicolumn{1}{c|}{LD}   &-31.4 &-42.5 &-43.5 &-37.5 &-29.1 &-33.5 &-36.2 \\\midrule
  \rowcolor{lightgray}\multicolumn{1}{c}{B-CANF~\cite{chen2024bcanf}}      & \multicolumn{1}{c}{TCSVT'23} & \multicolumn{1}{c|}{RA}  &-8.4 &10.6& 10.4 &30.4 &6.4& 8.7& 9.7\\
  \multicolumn{1}{c}{UCVC~\cite{yang2024ucvc}}      & \multicolumn{1}{c}{DCC'24} & \multicolumn{1}{c|}{RA}  &-17.7 &-29.1 &-24.4 &-17.1& -21.2 &-18.8 &-21.4\\
  \multicolumn{1}{c}{DCVC-B~\cite{sheng2025bi}}      & \multicolumn{1}{c}{TMM'25} & \multicolumn{1}{c|}{RA} &-25.2& -33.8 &-40.7 &-46.5 &-19.4 &-26.2 &-32.0\\
  \rowcolor{lightgray}\multicolumn{1}{c}{BiECVC}      & \multicolumn{1}{c}{Ours} & \multicolumn{1}{c|}{RA} &  \textcolor{purple}{$\bm{-33.8}$}   & \textcolor{purple}{$\bm{-42.8}$} & \textcolor{purple}{$\bm{-48.6}$}  & \textcolor{purple}{$\bm{-47.5}$} &    \textcolor{purple}{$\bm{-32.7}$} &  \textcolor{purple}{$\bm{-34.9}$}   & \textcolor{purple}{$\bm{-40.1}$}\\
   \bottomrule  
\end{tabular}}      
\end{threeparttable}}
\caption{{BD-Rate $(\%)$~\cite{bjontegaard2001calculation} comparison for MS-SSIM~\cite{wang2003multiscale}. The anchor is \textbf{DCVC-TCM}. \textbf{The Intra Period is $\bm{32}$ with $\bm{96}$ frames}.}}
\label{tab:rd_ssim}   
\end{table*}
\end{document}